\documentclass[conference]{sig-alternate-10pt}
\usepackage{balance}
\usepackage{pifont}
\usepackage{graphicx}
\usepackage{subfigure}

\usepackage{url}
\usepackage{xspace}
\usepackage{hyperref}

\hypersetup{
    colorlinks,
    citecolor=black,
    filecolor=black,
    linkcolor=black,
    urlcolor=black
}

\newenvironment{tightlist}{
\begin{list}{$\bullet$}{
   \setlength{\topsep}{0in} 
    \setlength{\partopsep}{0in}
    \setlength{\parskip}{0in}
    \setlength{\itemsep}{0in}
    \setlength{\parsep}{.2em}
    \setlength{\leftmargin}{1.0em}
    \setlength{\rightmargin}{0in}
    \setlength{\itemindent}{0in}
}}
{\end{list}}

\usepackage{booktabs,multirow}
\newcommand{\ra}[1]{\renewcommand{\arraystretch}{#1}}

\usepackage[linesnumbered,lined,ruled,commentsnumbered]{algorithm2e}

\newcommand{\sref}[1]{{\S\ref{#1}}}
\newcommand{\fref}[1]{\mbox{Figure~\ref{#1}}}
\newcommand{\tref}[1]{\mbox{Table~\ref{#1}}}
\newcommand{\eref}[1]{\mbox{Eq.~\eqref{#1}}}

\newcommand{\myparab}[1]{\smallskip\noindent\textbf{#1}\xspace}

\newcommand{\eg}{{\it e.g.}\xspace}
\newcommand{\ie}{{\it i.e.}\xspace}

\newcommand{\eat}[1]{}
\newcommand{\todo}[1]{\emph{TODO: #1}}

\newcommand{\scheme}{AMP\xspace}
\newcommand{\prob}{minimum window syndrome\xspace}
\newcommand{\cwnd}{\texttt{cwnd}\xspace}

\makeatletter
\def\@copyrightspace{\relax}
\makeatother

\begin{document}

\title{AMP: A Better Multipath TCP for Data Center Networks}

\numberofauthors{2}

\author{
Morteza Kheirkhah\\
\affaddr{University College London}\\
\email{m.kheirkhah@ucl.ac.uk}
\and 
Myungjin lee\\
\affaddr{University of Edinburgh}\\
\email{myungjin.lee@ed.ac.uk}
}


\maketitle

\pagestyle{plain}

\begin{abstract}
In recent years several multipath data transport mechanisms, such as MPTCP and
XMP, have been introduced to effectively exploit the path diversity of data
center networks (DCNs).
However, these multipath schemes have not been widely deployed in DCNs. We argue
that two key factors among others impeded their adoption: \emph{TCP incast} and
\emph{\prob}. First, these mechanisms are ill-suited for workloads with a
many-to-one communication pattern, commonly found in DCNs, causing frequent TCP
incast collapses.  Second, the syndrome we discover for the first time, results
in 2-5 times lower throughput for single-path flows than multipath flows, thus
severely violating network fairness.

To effectively tackle these problems, we propose \scheme: an adaptive multipath
congestion control mechanism that quickly detects the onset of these problems
and transforms its multipath flow into a single-path flow. Once these problems
disappear, \scheme safely reverses this transformation and continues its data
transmission via multiple paths. Our evaluation results under a diverse set of
scenarios in a fat-tree topology with realistic workloads demonstrate that
\scheme is robust to the TCP incast problem and improves network fairness
between multipath and single-path flows significantly with little performance
loss.
\end{abstract}
\section{Introduction} 
\label{sec:intro} 

Data center is a crucial infrastructure that drives the Internet today. A
large-scale data center comprises hundreds of thousands of servers, and hosts a
diverse set of online services that require high bandwidth, low latency or both
from the network. To meet such requirements, lots of recent
advances~\cite{dctcp, timely, mptcp, mmptcp, xmp, fuso, d2tcp, l2dct, tcprand,
  ictcp, dx, mtcp, tuning} have focused on improving TCP congestion control (CC)
algorithms, by leveraging path diversity~\cite{mptcp, mmptcp, xmp, fuso},
exploiting explicit congestion signals from network switches~\cite{dctcp, xmp,
  d2tcp, tuning, l2dct}, measuring delays~\cite{dx, timely}, etc.

In this paper we focus on striking a right balance between throughput and
latency at transport layer. To that end, one seemingly natural way is to
combine a multipath tranport protocol (\eg, MPTCP~\cite{mptcp}) and a
low-latency tranport protocol (\eg, DCTCP~\cite{dctcp}) in that the former
usually achieves a high throughput and the latter keeps switch buffer occupancy
low by exploiting Explicit Congestion Notification (ECN). Thus, the crux of this
idea is to maintain multiple subflows per connection using a multipath mechanism
and each subflow runs a low-latency protocol such as DCTCP.

This makes sense because any single transport protocol is difficult to meet
high-throughput and low-latency requirements. For example, Equal Cost Multi Path
(ECMP) routing would be likely to cause collisons among long-lived DCTCP flows
on the same link, which can degrade throughput substantially.
In contrast, MPTCP is good at fast load-balancing, overcoming the shortcoming of
ECMP. However, it tends to occupy switch buffers aggressively, thus hurting the
performance of latency-sensitive short flows.

We conduct such integration, which we call Data Center MultiPath (DCM, in
short), and examine a similar existing approach called XMP~\cite{xmp}. We find
that both schemes can provide fast load-balancing while keeping switch buffer
occupancy low. However, two major challenges ---\emph{TCP incast} and
\emph{\prob}--- still render these approaches less practical as a transport
protocol for DCNs.

We carefully examine these problems when multipath schemes are in use
(\sref{s:motive}). Multiple subflows in these MPTCP variants boost the
possibility of TCP incast in a many-to-one communication pattern while for
example senders and a receiver are co-located in a single rack. Worse, in that
setting, network resource competition between ECN-capable multipath (\eg, DCM)
and single-path (\eg, DCTCP) flows causes a serious co-existence problem, which
we name \emph{\prob}. Surprisingly, the syndrome consistently makes multipath
flows achieve 2-5$\times$ more throughput than single-path flows, thus severely
violating fairness among the TCP flows. We find out that using both multiple
subflows and small ECN marking threshold is behind the syndrome.

Finally we propose \scheme, an adaptive multipath congestion control algorithm
that is robust to the TCP incast problem and effectively handles the \prob
with little performance compromise on both throughput and latency. In addition
to good fairness and high performance, we design \scheme such that it is simple
enough to keep its behavior traceable and its overheads low, and can shift
traffic quickly from congested paths to less congested paths. \scheme requires
none of sophisticated mechanisms such as RTT-dependent congestion window (\cwnd)
increase (in MPTCP) and dynamic \cwnd decrease (in DCTCP).

\scheme's approach is simple but effective: it simply transforms a multipath
flow into a single-path flow at the onset of the problems. The key in \scheme is
the early detection of the problems. We leverage the fact that all subflows of a
multipath flow have the smallest congestion window value, which is a good
indicator that all of the subflows compete with other flows on a single link. If
the minimum window state across all subflows remains for a small time period
(\eg, 1-3 RTTs), \scheme executes this transformation by deactivating all
subflows but one. If \scheme no longer receives ECN-marked packets for some time
period (\eg, 8 RTTs), it reactivates all suspended subflows (\sref{s:design}).
Our evaluation shows that this neat technique
substantially mitigates TCP incast and improve fairness without any side-effect
(\sref{s:micro}).

\scheme also simplifies congestion control operations, which keeps \scheme
easily traceable and its overheads low. \scheme just increases only one window
per RTT across all subflows, similar to the behavior of a single-path TCP
whereas the other schemes consider RTTs of all subflows to update their \cwnd.
In response to ECN signals, \scheme cuts \cwnd by a constant factor instead of
dynamically adjusting it based on the fraction of marked packets
(\sref{s:algo}). Our extensive evaluations in a large-scale fat-tree topology
with realistic traffic matrix demonstrate that \scheme under incast-like
workloads works better than and in other workloads performs as well as the
existing solutions, despite its simplicity (\sref{s:macro}).

Overall, this paper makes three main contributions:

\begin{tightlist}
\item To the best of our knowledge, we report for the first time that the \prob
  can do exist when ECN-capable multipath and single-path TCPs are deployed in
  data centers. We carefully examine its root cause.

\item We propose \scheme\footnote{The \scheme source code is available at
    \url{https://github.com/mkheirkhah/amp}. Note that we have implemented \scheme on top of our custom implementation of MPTCP in Network Simulator-3 (NS-3)~\cite{Morteza}.}, an adaptive multipath TCP for
  data center networks that effectively copes with the TCP incast and
  \prob. \scheme is resilient against the incast problem and ensures graceful
  co-existence with single-path TCP flows.


\item We evaluate \scheme over a wide variety of scenarios in a
  large-scale fat-tree topology, and demonstrate that \scheme
  mitigates buffer inflation and achieves higher fairness and
  comparable performance against existing multipath protocols. 
\end{tightlist}

\section{Preliminary}
\label{s:prelim}

In this section we review two multipath mechanisms to facilitate our later
discussions: (1) DCM, a new extension of MPTCP that combines the congestion
control of MPTCP and DCTCP together;
and (2) XMP, an existing proposal.

\subsection{DCM}
\label{s:dcm}
An intuitive and reasonable approach is to combine MPTCP and DCTCP. The main
idea is to enable each subflow of MPTCP with the ECN response mechanism of
DCTCP.
On top of the basic MPTCP algorithm,
which swiftly shifts traffic from highly congested to less congested paths, DCM
handles ECN-marked packets similar to DCTCP for each subflow. That is, each subflow of DCM adjusts its sending rate in proportion to the extent of congestion, represented by the amount of ECN-marked packets. For instance, when a subflow rapidly reduces its \cwnd due to receiving a large amount of ECN-marked packets over a few windows of data, DCM moves the traffic from that subflow to other subflows with better network condition (\eg with larger \cwnd and low RTT). In this way, each subflow of DCM follows DCTCP to reduce its \cwnd and MPTCP to increase its \cwnd.

\noindent \textbf{The DCM does, in short:}
\begin{tightlist}
\item For each ACK
  on subflow $s$, $w_{s} \leftarrow w_{s} + \min(\frac{a}{w_{total}},\frac{1}{w_{s}})$
\item For each loss, $w_{s} \leftarrow \frac{w_{s}}{2}$
\item For first marked ACK in a window, $w_s \leftarrow w_s (1-\frac{\alpha_s}{2})$
\end{tightlist}

$w_s$ is a \cwnd size of subflow $s$, $w_{total}$ is $\sum_{r} w_{r}$ for all
$r$ and $a$ controls the aggressiveness of \cwnd increase across all subflows.
The following formula calculates the value of $a$:
\begin{equation} \label{eq:mptcp}
a = w_{total} \frac{\max_{r}\ (w_{r}/rtt_{r}^2)} {(\sum_{r} (w_{r}/rtt_{r}))^2}
\end{equation}

Here $\max_{r}$ is the maximum value across all subflows. For instance, when an MPTCP flow encounters a path with high RTT and low packet drop probability, it increases its aggressiveness to fully utilize that path. The aggressiveness is also capped by $\frac{1}{w_{s}}$ to prevent a subflow to increase its \cwnd more than one segment per RTT, ensuring that a subflow, and in turn, the MPTCP flow, is not harming other competing (possibly, single-path) flows. 

$\alpha_s$ is an estimate of the fraction of marked packets on subflow $s$ and is updated once per window of data (roughly an RTT) as follows:
\begin{equation} \label{eq:dcm}
\alpha_s = (1 - g)\alpha_s + gF_s
\end{equation}

$F_s$ is the fraction of marked packets (in the last window of data) on subflow $s$; $g$ is a (constant) weight coefficient for
exponentially averaging $\alpha_s$. 
When $\alpha_s\rightarrow0$, $w_s$ decreases gently; as $\alpha_s\rightarrow1$, $w_s$
does more aggressively. 

\subsection{XMP}
\label{s:xmp}
XMP is another multipath congestion control algorithm that aims to strike a
balance between latency-throughput trade-offs. XMP combines an ECN-based scheme
for controlling the buffer occupancy in switches and a rate-based congestion
control algorithm for balancing traffic among its subflows.

\noindent \textbf{The XMP does, in short:}
\begin{tightlist}
\item Every window of data on subflow $s$, $w_{s} \leftarrow w_{s} + \delta_{s}$
\item For each loss, $w_{s} \leftarrow \frac{w_{s}}{2}$
\item For first marked ACK in a window, $w_{s} \leftarrow w_{s} (1-\frac {1}{\beta})$
\end{tightlist}

$\delta_{s}$ dictates the amount of \cwnd increase for each subflow, calculated once per window of data; and $\beta$ is a fixed reduction factor (set to 4 in \cite{xmp}). The value for $\delta_{s}$ is calculated by the following formula:
\begin{equation} \label{eq:xmp}
\delta_{s} = \frac{rtt_{s}}{rtt_{min}} \times \frac{w_{s}/rtt_{s}} {\sum_{r} (w_{r}/rtt_{r})} 
\end{equation}

XMP is in principle similar to MPTCP and DCM, but there are differences, too.
One of them is that in XMP, network congestion is signaled via packet queuing
delay (inferred through RTT) and ECN-marked packets.

\begin{figure}[t]
\centering 
\includegraphics[width=0.33\textwidth]{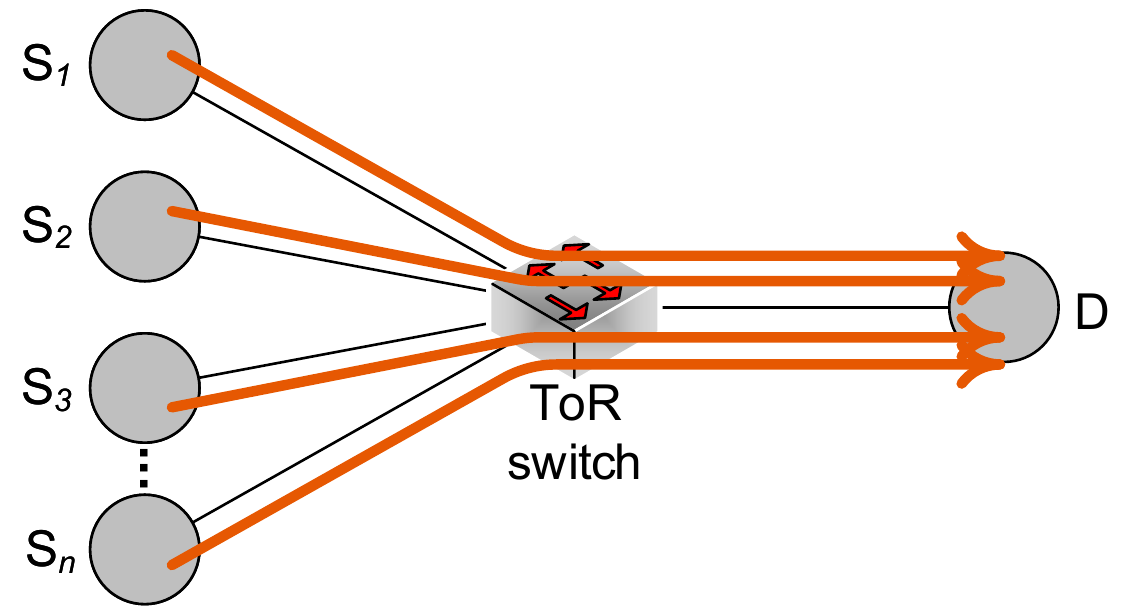}
\vspace{-.1in}
\caption{A many-to-one communication scenario over a 10Gbps bottleneck link.}
\vspace{-.1in}
\label{fig:startopo}
\end{figure}

\section{Issues of MPTCP variants}
\label{s:motive}

MPTCP and its ECN-capable variants have not been widely deployed in DCNs. While
there may be several other reasons, we identify two key technical issues. First,
the ECN-capable MPTCP variants (DCM and XMP) are unable to handle incast-like
traffic; many applications (MapReduce~\cite{mapreduce},
Partition/Aggregate~\cite{dctcp}, etc.) have a many-to-one communication pattern
that is prevailing in DCNs. Second, the ECN-capable MPTCP variants fail to
gracefully coexist with an ECN-capable single-path TCP such as DCTCP; an MPTCP
variant can hurt DCTCP flows' throughput significantly.  We call this problem
the \prob. In what follows, we demonstrate the impact of these two problems via
simulation under a simple topology shown in \fref{fig:startopo}.

\begin{figure}[t]
\centering 
\includegraphics[width=0.4\textwidth]{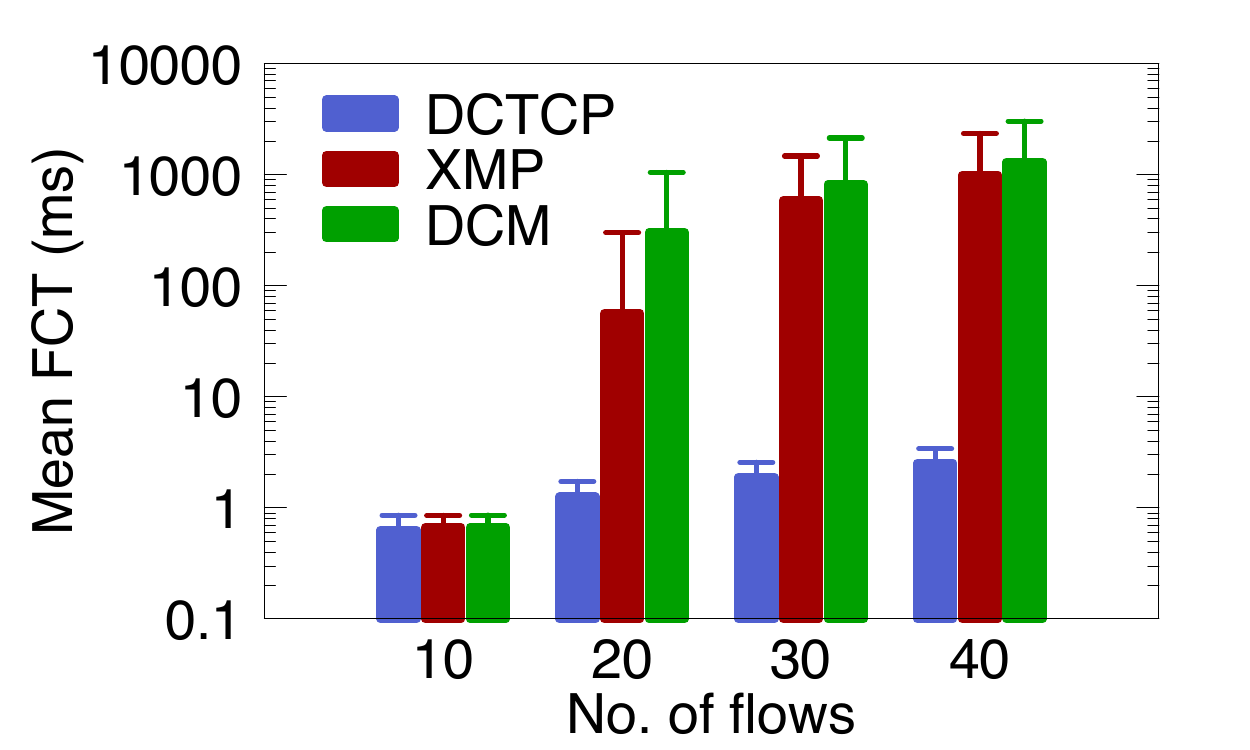}
\vspace{-.1in}
\caption{Impact of the TCP incast on different multipath protocols (DCM and XMP)
  and DCTCP. DCM and XMP use four subflows per connection. File size is 128KB,
  link capacity is 10Gbps, and switch buffer size is 100 packets. The y-axis is
  log-scaled.}
\vspace{-.1in}
\label{fig:incast128kb} 
\end{figure}

\begin{figure*}[t]
\centering 
\subfigure[Normal situation]{
\includegraphics[height=0.13\textheight]{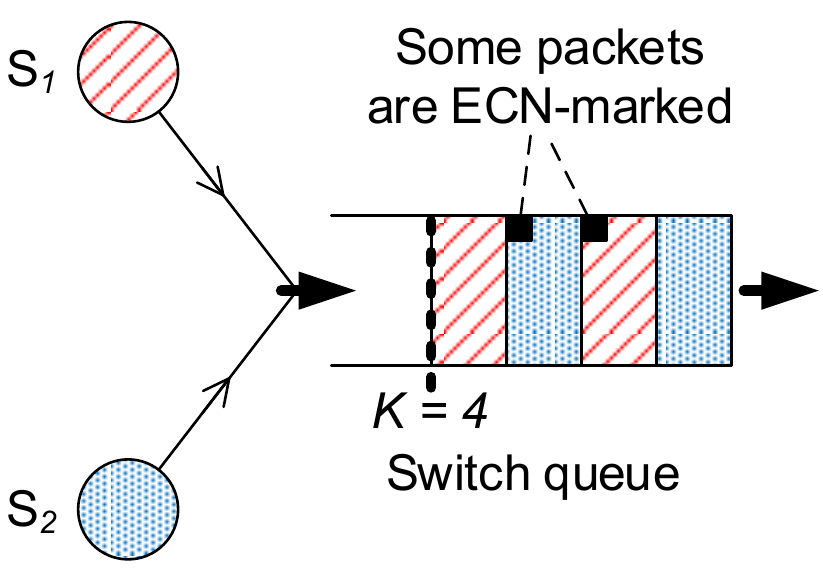}
\label{fig:sp2flows}}
\hspace{.1in}
\subfigure[Persistent buffer inflation]{
\includegraphics[height=0.13\textheight]{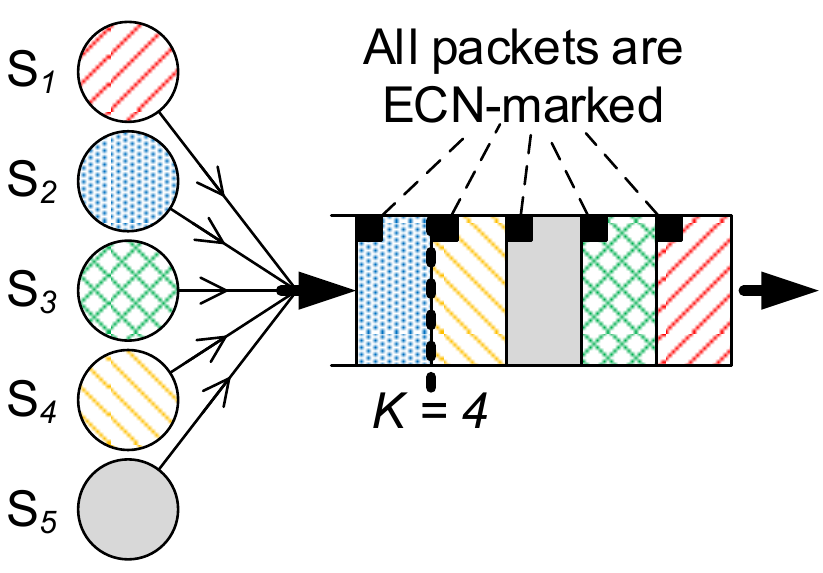}
\label{fig:sp5flows}}
\hspace{.1in}
\subfigure[Minimum window syndrome]{
\includegraphics[height=0.13\textheight]{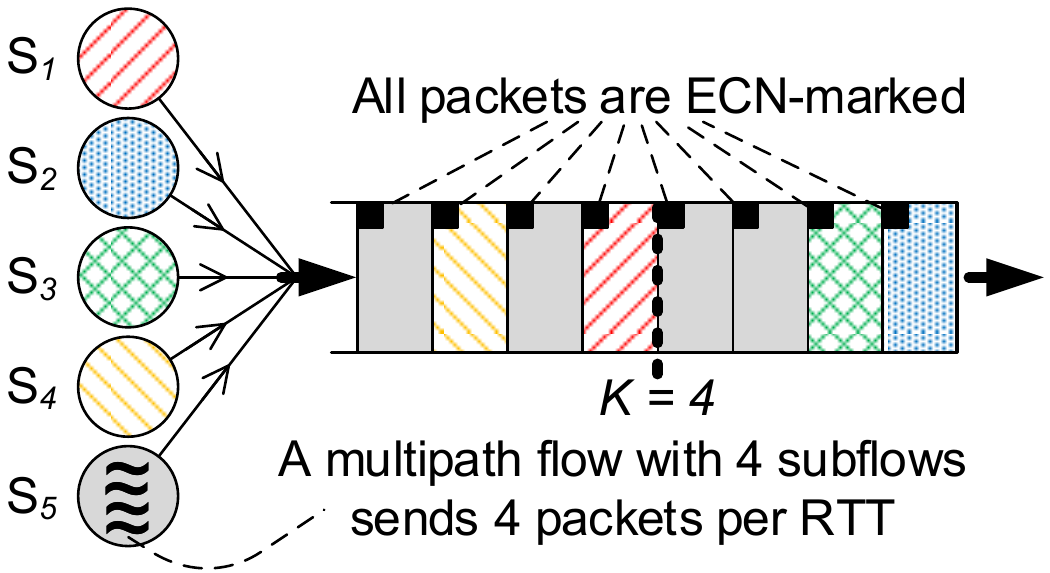} 
\label{fig:spmpflows}}
\vspace{-.1in}
\caption{Illustration of the \prob. The syndrome leads to severe unfairness and
  escalates the likelihood and impact of persistent buffer inflation
  significantly.}
\vspace{-.1in}
\label{fig:exmws} 
\end{figure*}

\subsection{TCP incast}
\label{s:incast_motive}

TCP incast can happen in applications which have barrier-synchronized workload
and a high fan-in communication pattern. A unique characteristic of the
barrier-synchronized workload is that traffic of multiple TCP flows to the same
destination arrives in a bursty fashion at a bottleneck switch, which has a
shallow buffer. This causes bursty packet losses and eventually triggers
expensive timeout at TCP senders, which substantially delays the completion of a
job~\cite{dctcp}. 

TCP incast is a well-studied topic~\cite{dctcp, ictcp} and for instance DCTCP
mitigates the problem using ECN. Unfortunately, the ECN-capable MPTCP variants
are still susceptible to the TCP incast even in the help of ECN. To demonstrate
that, we create a simulation environment as shown in \fref{fig:startopo} using
NS-3~\cite{ns3}. The simulation setup is as follows. Every 1 second $k$ number
of multipath flows join to a bottleneck link with a fixed interval of $50\mu s$
where $k = 10, 20, 30, 40$ while setting the flow size to 128KB. Each simulation
lasts for 20 sec. Each multipath flow has 4 subflows. In the setup, we test DCM
and XMP. We also separately run DCTCP as baseline.

\fref{fig:incast128kb} shows that DCTCP overall outperforms DCM and XMP. In many
cases the average flow completion time (FCT) of DCTCP is almost 1-2 orders of
magnitude shorter than that of DCM and XMP; when $k = 30$, the average FCT of
DCTCP is about 2ms whereas that of DCM and XMP is over 800ms. Furthermore, the
FCT distribution of DCTCP has a narrow standard deviation (\ie, the whisker bar
in the graph), but the standard deviation of XMP and DCM is large (less than 1
millisecond for DCTCP vs. above 1 second for DCM and XMP). This means that the
other schemes have a long-tailed FCT distribution and make some flows experience
much higher FCTs (due to retransmission timeouts).

From these results, it is evident that the multipath variants cannot handle the
TCP incast problem. The reason is somewhat obvious. The MPTCP variants maintain
4 subflows. Hence, one multipath flow generates at least 4 packets per
RTT. More number of multipath flows implies a sharp increase of the probability
of burst packet losses. For example, every RTT 30 multipath flows shown in
\fref{fig:incast128kb} generate at least 120 packets, which are far exceeding
the queue length (\ie, 100 packets in this case) of the bottleneck switch.

Without giving up the benefit of a multipath protocol, a (practical and possibly
natural) way to deal with this problem may be allowing both multipath and
single-path protocols and letting them share the DCN fabric. The basic idea is
to permit DCTCP for latency-sensitive applications and multipath protocols for
bandwidth-hungry services. However, keeping graceful co-existence of these two
different protocols turns out to be challenging, which we discuss next.

\subsection{Minimum window syndrome}

In the presence of ECN-capable multipath (\eg, DCM and XMP) and single-path TCP
flows (\eg, DCTCP), serious unfairness between them can occur. The key
characteristic of this problem is that when all the subflows of DCM or XMP flows
compete with DCTCP flows on the same bottleneck link, DCTCP flows obtain 2-5
times less amount of bandwidth and higher queueing delay than they should. We
call this co-existence problem the \prob.

\myparab{Triggering the syndrome.} We discuss when the syndrome can occur
through examples shown in \fref{fig:exmws}. We assume that network switches
ECN-mark packets only if their instant queue length is larger than a marking
threshold $K$. Such switches are widely deployed in data centers. To keep the
discussion simple, let us assume $K = 4$ and zero propagation delay. That is, as
soon as packet leaves the queue, sender can send a new packet as it receives
acknowledgment instantly.

In \fref{fig:sp2flows}, two single-path flows share the bottleneck link fairly
by generating on average two packets per RTT (bounded by queuing delay); in
other words, \cwnd in each flow oscillates between 1-3 packets.

Now suppose that 5 single-path flows compete with each other as illustrated in
\fref{fig:sp5flows}. Because $K = 4$, a new arriving packet finds the queue
length is always equal to $K$, meaning that it is the 5th packet in the
queue. Thus, all packets across flows are ECN-marked all the time, and each flow
is forced to reduce its \cwnd to one packet. Even though there is no way to
further reduce \cwnd (as it is one packet), the number of packets arriving at
the queue always exceeds $K$. This causes \emph{persistent buffer inflation}
(also discussed in \cite{alizadeh11}), but there is no unfairness across
flows.

Finally, \fref{fig:spmpflows} illustrates a case where the single-path flow in
$S_5$ is replaced with one multipath flow having four subflows. Similar to the
previous case, all packets across flows are ECN-marked. Even if the \cwnd of the
single-path flows and all subflows reduces to one packet, the number of packets
in flight far exceeds $K$ all the time. However, since all the subflows belong
to one multipath flow, the flow ends up sending four times more packets than
single-path ones. Furthermore, the syndrome substantially escalates the
likelihood and impact of the persistent buffer inflation (see the buffer length
twice as large as $K$ in \fref{fig:spmpflows}), which can potentially harm
latency-sensitive short flows.

\myparab{Conditions for the syndrome.} In reality, BDP needs to be considered
and is a few tens of packets in DCNs~\cite{xmp, tuning}. Thus, to create the
syndrome, more than ($BDP + K$) number of flows are necessary. However, the
MPTCP variants set \emph{minimum congestion window size} ($cwnd_{min}$)---an
internal constant that governs the minimum number of packets a sender can send
regardless of congestion level---to two packets\footnote{MPTCP and XMP use two
  packets to probe congestion level on each path (see a detailed discussion in
  \cite{damon_cc} and Algorithm~1 in \cite{xmp}). DCTCP also uses two packets
  originally, but a recent study proposed to use one packet for the value (see
  page 11 in \cite{judd}) and the DCTCP source was patched accordingly. Unless
  otherwise stated, we set $cwnd_{min} = 2$ for consistency in this paper.}.
Thus, the number of flows including single-path flows and subflows in multipath
flows should be larger than $(BDP + K) / cwnd_{min}$.

\begin{figure}[t]
\centering 
\subfigure[{
($K$, $r$) = ($10$, \{$2 \ldots 8$\}). Average goodput of 8 DCTCP flows and  one
XMP flow.}]{
\includegraphics[width=0.3\textwidth]{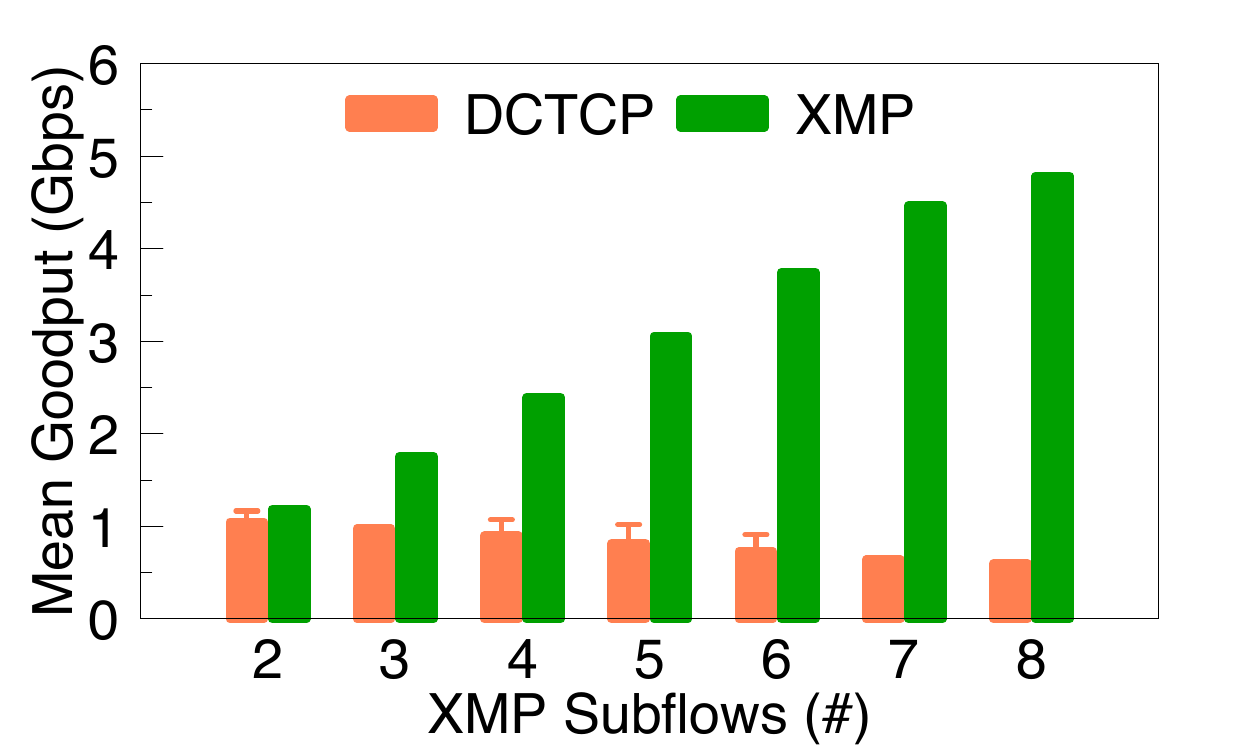}
\label{fig:mwssubflows}}
\subfigure[
($K$, $r$) = ($20$, $4$). \mbox{Average} goodput of varying number of DCTCP
flows and one XMP flow.]{
\includegraphics[width=0.3\textwidth]{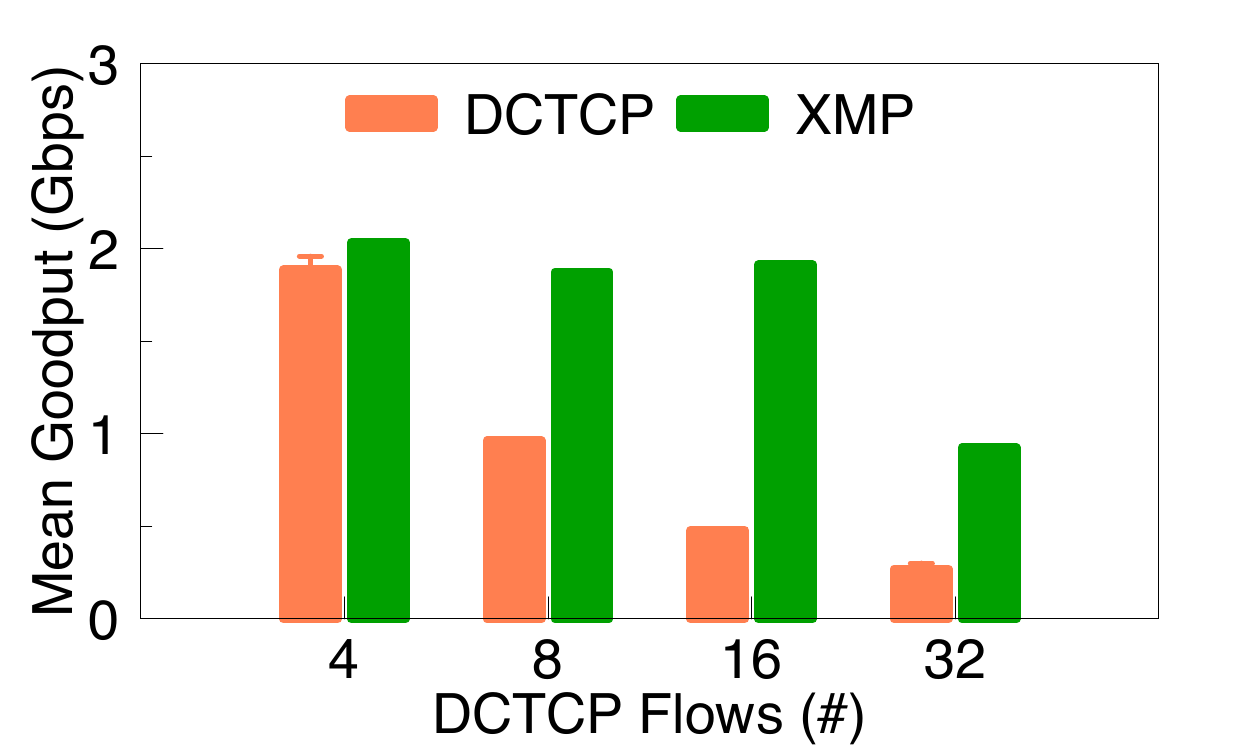}
\label{fig:mwsk20}}
\vspace{-.1in}
\caption{The \prob under various conditions. $K$: ECN marking threshold,
  and $r$: the number of subflows.}
\vspace{-.1in}
\label{fig:motive} 
\end{figure}

\myparab{Demonstration of the syndrome.} Now suppose a setup in
\fref{fig:startopo} where an ECN-enabled switch connects $n$ sending servers and
one receiving server. The receiver is equipped with DCTCP, DCM and XMP;
server $S_1$ runs DCM (or XMP) having $r$ subflows, and the remaining $n-1$
servers with DCTCP ($n \ge 2$). Those $n$ senders send traffic to the receiver.

We do various simulations by varying parameters and study the impacts of the
syndrome. We change the simulation duration from 10ms to 1 sec, use 1Gbps and
10Gbps link and test both DCM and XMP. Across these variations, we observe a
very similar trend. Thus we only show the results of the 1 sec duration over
10Gbps link using XMP in interest of space. We depict a setting as ($K$, $r$)
where $K$ is the ECN marking threshold and $r$ is the number of subflows.

\emph{Varying number of subflows:} Given 8 DCTCP flows and one DCM or XMP flow,
we vary the number of subflows from 2 to 8, while setting $K=10$ and
$cwnd_{min}=2$ as suggested in \cite{xmp}; thus, the setting is ($10$, \{$2
\ldots 8$\}).

\fref{fig:mwssubflows} shows that the syndrome begins as soon as the DCM or XMP
flow starts to use three subflows or more. When four subflows are used, the XMP
flow obtains 2.3$\times$ higher goodput than DCTCP flows. The figure clearly
demonstrates that the number of subfows is a key factor that triggers the
problem.
DCTCP flows seem to have no problem in the 2-subflow case. However, the problem
recurs when at least about 16 DCTCP flows are in use (not shown for brevity).
Worse, using two subflows costs about 10\% goodput loss (\eg, 1Gbps out of
10Gbps rate) when compared to using four subflows~\cite{xmp}. Also, a number of
subflows (\eg, 8 subflows) are in general beneficial when there exist lots of
parallel paths in a large DCN~\cite{mptcp}. Thus, using a smaller number of
subflows is not a fundamental solution.

\emph{Different marking threshold:} As a small marking threshold can be a
potential cause of the problem, increasing $K$ may be useful.
However, this can also introduce an additional delay,
which might hurt the flow completion time of latency-sensitive short
flows. Nevertheless, we test $K = 20$.
With the setting ($20$, $4$), we vary the number of DCTCP flows.

\fref{fig:mwsk20} shows that increasing the marking threshold marginally
alleviates the problem; given 8 DCTCP flows, a goodput gap between DCTCP and DCM
or XMP is a factor of two. In contrast, recall that the gap is a factor of 2.3
under the same condition in \fref{fig:mwssubflows}. We also tested a case where
$cwnd_{min} = 1$ while keeping the setting as ($10$, $4$). This reduced the
likelihood of the syndrome, but we observed that a slight increase of the number
of DCM or XMP flows (from 1 to 4) triggered the syndrome, when 8 DCTCP flows are
given (the exact graph is omitted).

\eat{
\emph{Different minimum congestion window size:} We test a case where
$cwnd_{min} = 1$ while leaving other parameters unchanged from the default
setting. Hence, the setting is ($10$, $1$, $4$). From \fref{fig:mwscwnd1}, we
find reducing $cwnd_{min}$ effective because it requires more number of DCTCP
flows (16 from the figure) to trigger the problem. However, notice that there is
only one DCM or XMP flow in this case. When we increase the number of DCM or XMP
flows from 1 to 4, again 8 DCTCP flows are sufficient to trigger the syndrome
(the exact graph is omitted).
}

\myparab{Summary.} We obtain two key findings from these results. First, the
condition that triggers the \prob is relatively simple: the total number of
packets in flight from both multipath and single-path flows should exceed BDP
plus $K$ frequently. In our setup, BDP is 20 packets. In \fref{fig:mwssubflows},
the condition begins to hold when the setting has 3-4 subflows for the XMP flow
and 8 DCTCP flows (the average number of packets in flight is about 30-32).
Second, tweaking those parameters either alleviates the problem marginally or
makes performance loss inevitable.

\section{Design}
\label{s:design}

We propose \scheme, a multipath congestion control mechanism that coexists well
with ECN-capable single-path TCPs and is resilient against TCP incast. In
designing \scheme, in addition to the obvious objectives---high throughput and
low latency, we have the following design objectives:

\begin{itemize}
\item \emph{Good fairness:} Multipath and single-path TCP flows should be able
  to achieve their fair share of bandwidth at a bottleneck link, even in the
  presence of an incast-like traffic pattern. 
\item \emph{Fast traffic shifting:} Multipath flows should be able to avoid
  congested paths quickly. This especially helps latency-sensitive short
  single-path TCP flows experience less impact due to congestion.
\item \emph{Simplicity:} An algorithm should be kept as simple as possible so
  that its behaviors are easily analyzed and its overheads are kept low.
\end{itemize}

To achieve the above objectives, we deliberately test existing solutions: DCM
and XMP.
In analyzing them, we make several key observations essential for our design.

\subsection{Key observations}
\label{s:observe}

\myparab{(1) The number of subflows for a multipath flow should not be static.}
Multiple subflows are in general beneficial to obtain high throughput. On the
contrary, in the presence of the TCP incast and \prob,
it is effective to have a smaller number of subflows (ideally, one subflow), as
discussed in \sref{s:motive}. However this costs throughput performance. Thus
having the static number of subflows can only achieve either good fairness
against single-path flows or high throughput, but not both of them. Thus, the
number of subflows should be adjusted adaptively, which can be done by
(de)activating subflows in an online fashion. However, it is inappropriate to
deactivate subflows incrementally because mitigating the two problems can take
too long, which may cause significant queuing delay to latency-sensitive short
flows over a longer period of time.

\myparab{(2) The \cwnd values in subflows are a cue for the TCP incast and
  syndrome.} Detecting these problems early is key to adjusting the number of
subflows. We notice that when these problems are about to occur, subflows are in
a unique status where the \cwnd values across all subflows are always equal to a
minimum (\eg, two packets in \cite{xmp, mptcp}). This is a good indicator that
these problems are in effect because it is unlikely that all subflows of a
multipath flow passing through different paths face excessive congestion,
especially in a large-scale data center that has 100s of parallel paths between
a pair of source and destination.

\begin{figure}[t]
\centering
\includegraphics[width=.4\textwidth]{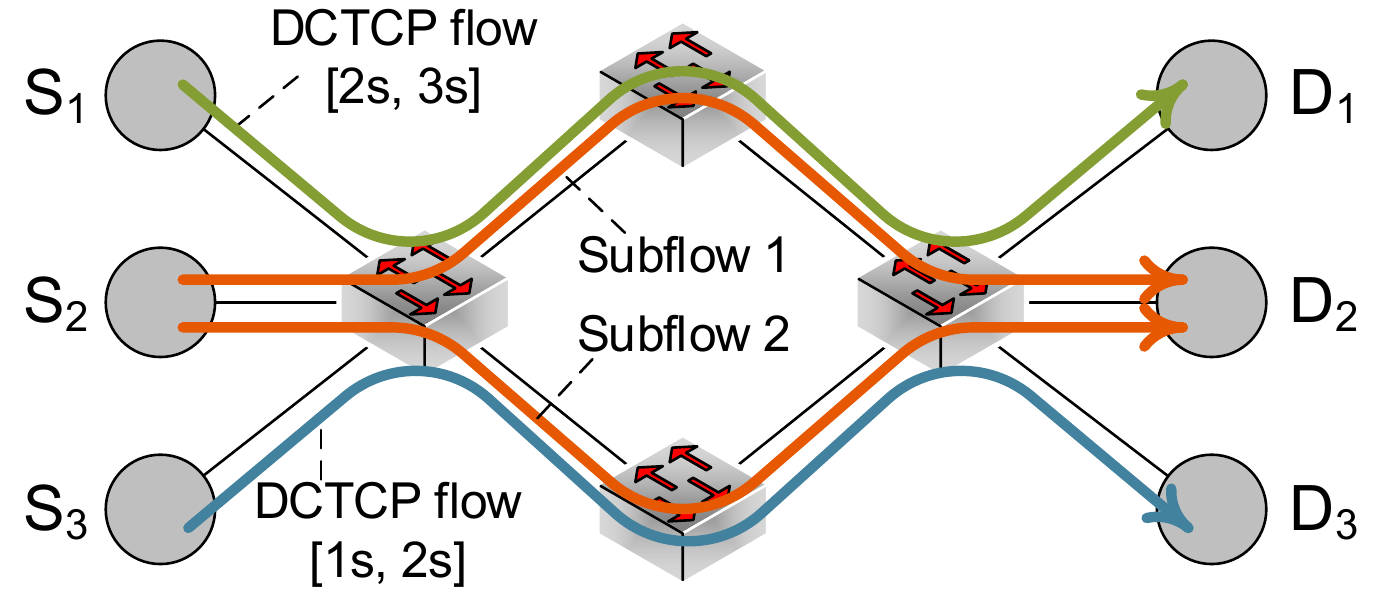} 
\vspace{-.1in}
\caption{A setup for testing traffic shifting time. An orange line
  represents a subflow of a multipath flow.}
\vspace{-.15in}
\label{fig:expts}
\end{figure}

\myparab{(3) Adaptive cutback of \cwnd at subflow slows down traffic shifting.}
One of differences between DCM and XMP is the response mechanism to ECN-marked
packets. In DCM a subflow cuts its \cwnd in proportion to the fraction of marked
packets over a window (see \eref{eq:dcm}); whereas in XMP a subflow decreases
its window by a constant factor $\beta$ (see \sref{s:xmp}). To understand the
effect of this difference, we modify MPTCP to reduce the \cwnd of subflow by a
constant factor (we use $\beta=4$) when it sees ECN-marked packets and examine
traffic shifting times for MPTCP with $\beta=4$ and DCM.

\begin{figure}[t]
\centering
\subfigure[MPTCP ($\beta = 4$)]{
\includegraphics[width=0.22\textwidth]{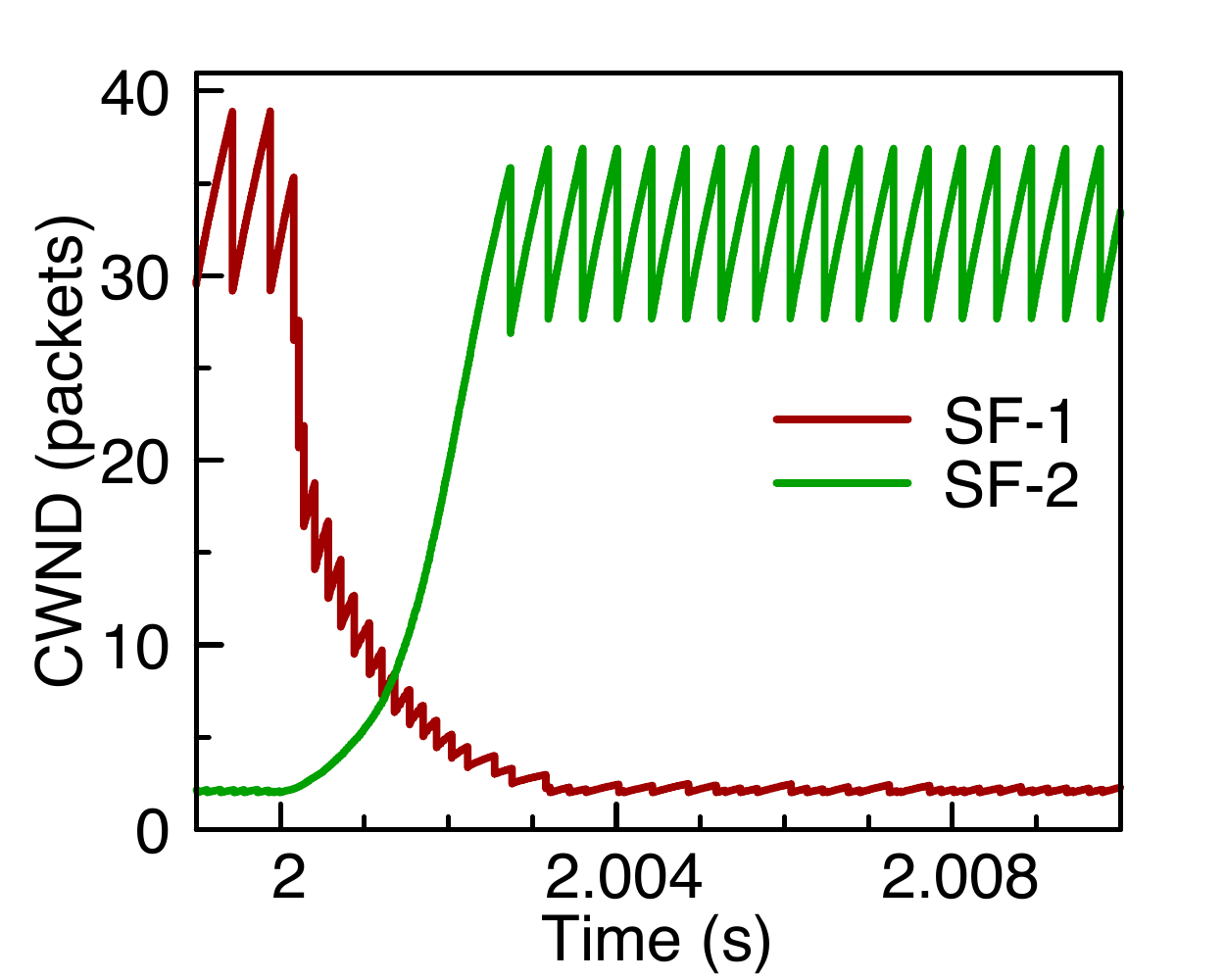}
\label{fig:trash_ecn}}
\subfigure[DCM]{
\includegraphics[width=0.22\textwidth]{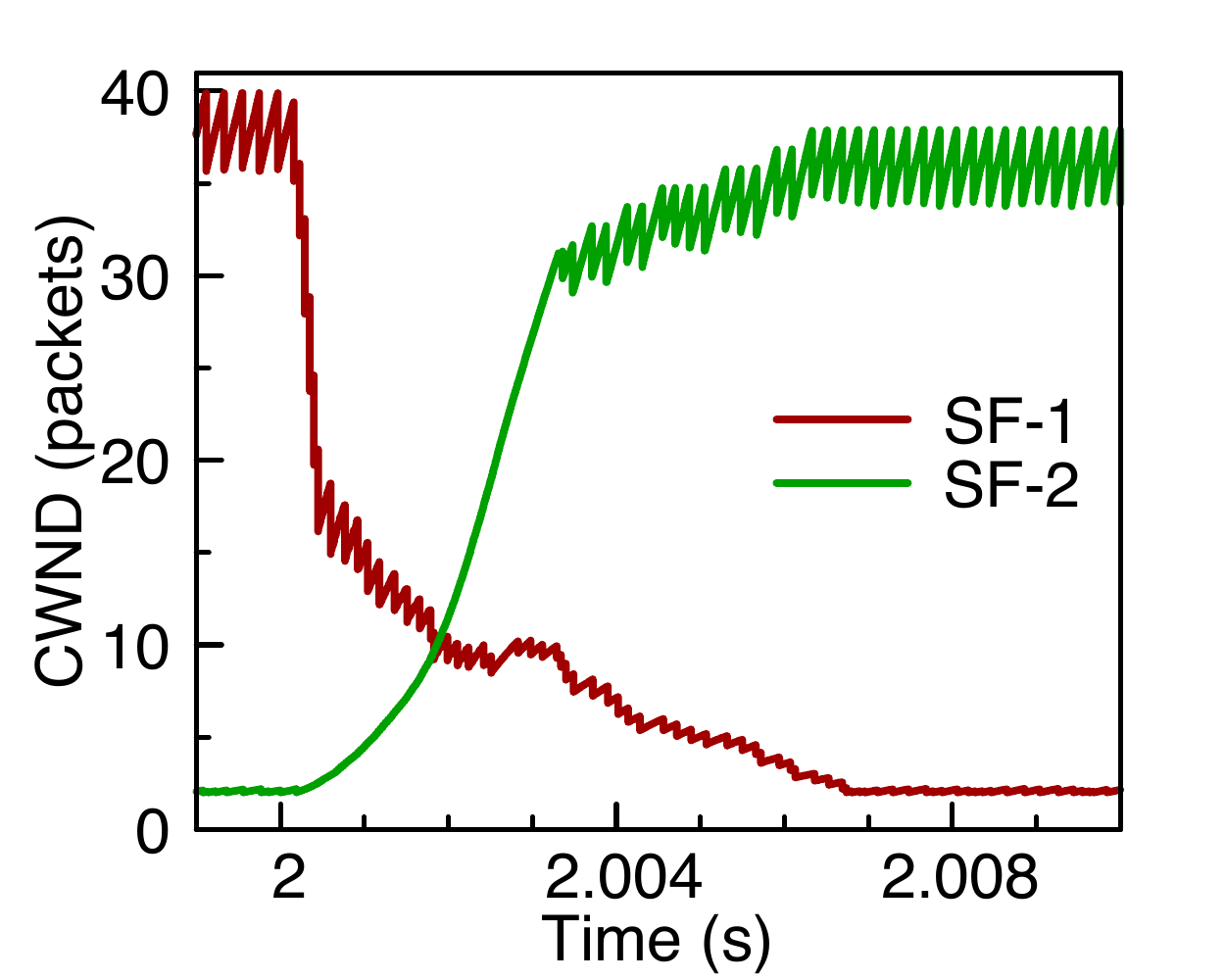}
\label{fig:trash_dctcp}}
\vspace{-.1in}
\caption{Traffic shifting times of MPTCP and DCM. MPTCP finishes its traffic
  shifting at 2.003s and DCM does at 2.007s; DCM is 4ms slower than MPTCP.}
\vspace{-.1in}
\label{fig:trashtimes}
\end{figure}

Given a topology shown in \fref{fig:expts}, a multipath flow (DCM or MPTCP) with
two subflows begins to traverse from $S_2$ to $D_2$ at 0s. Then, $S_3$ sends
traffic to $D_3$ using DCTCP within interval (1s, 2s) and another DCTCP flow
from $S_1$ to $D_1$ for (2s, 3s). At 2s, a multipath flow is sending its entire
traffic through the upper path and we plot how \cwnd of each subflow varies
within interval (1.999s, 2.01s) after the second DCTCP flow appears on the
upper path. \fref{fig:trashtimes} shows that MPTCP finishes traffic shifting
about 4ms faster than DCM.

The reason is because a DCTCP subflow in DCM conservatively reduces \cwnd based
on the fraction of marked packets. Hence, even if there exists a congestion-free
path, a DCM flow shifts its traffic slowly. In contrast, with a constant factor
(\eg, $\beta = 4$), MPTCP is aggressive enough to make a subflow on the
congested path quickly reduce its window, thereby achieving faster traffic
shifting than DCM.  This conservative nature of DCTCP perfectly makes sense if a
flow traverses one path only. However, because the subflows of a multipath flow
travel through multiple different paths in general, it is more appropriate to
get rid of traffic from the congested path rather than to withstand against
congestion.

\myparab{(4) RTT measurements of subflows are unnecessary for updating their
  \cwnd.} Interestingly, both DCM and XMP rely on RTT measurements in
increasing \cwnd of subflows. DCM inherits MPTCP's design principles, one of
which targets to address the RTT mismatch issue~\cite{damon_cc} that can occur
when there are paths with high RTT and low loss probability and paths with low
RTT and high loss probability.
However, higher RTT typically means large queuing delay and hence high loss
probability in DCNs because DCNs usually have a symmetrical structure where all
paths between a pair of servers have the same length. Thus, DCNs have no paths
that cause the RTT mismatch problem.

Moreover, ECN tends to equalize RTTs throughout the data center network when
network switches react to instant queue length with a small ECN marking
threshold~\cite{dctcp, xmp}. Assuming 5-hop paths with 10Gbps links, 10 packets
of marking threshold and 1500B packets, a maximum RTT difference is just about
108$\mu s$. In average cases, as the utilization of network links increases, the
RTT difference will become even smaller. Thus, differentiating the sending rate
of each subflow based on such a small RTT difference would not bring much
benefit. Even in a case that a path is highly congested,
sources can quickly identify it with ECN signals and do traffic shifting
accordingly.

\begin{algorithm}[t]
  \DontPrintSemicolon
  \tcc{Subflow suppression/release}
  \SetKwProg{Fn}{}{}{}\SetKwFunction{SS}{SuppressSubflows}%
  \Fn(){\SS{nRound}}{
    nSF = 0 \tcc*[f]{counter for subflows}\; \label{alg:supcheck_begin}
    \For{subflow $s \in [1, \ldots, n]$}{
      \lIf{$w_s$ = $cwnd_{min}$}{nSF $\leftarrow$ nSF + 1}
    }
    \lIf{nSF = $n$}{nRound $\leftarrow$ nRound + 1}
    \lElse{nRound $\leftarrow$ 0}  
     \lIf{nRound < $\gamma$}{\Return} \label{alg:supcheck_end}
    \For{subflow $s \in [2, \ldots, n]$ \tcc{at $\gamma$ rounds}}{ \label{alg:sup_begin}
      $active_s$ $\leftarrow$ false\; 
    }\label{alg:sup_end}
  }
  \SetKwProg{Fn}{}{}{}\SetKwFunction{REL}{ReleaseSubflows}%
  \Fn(){\REL{ACK, nRound}}{
    \lIf{ACK.marked}{nRound $\leftarrow$ 0} \label{alg:relcheck_begin}
    \lElse{nRound $\leftarrow$ nRound + 1}
    \lIf{nRound < $\tau$}{\Return} \label{alg:relcheck_end}
    \For{subflow $s \in [2,\ldots,n]$ \tcc{at $\tau$ rounds}}{ \label{alg:rel_begin}
      $active_s$ $\leftarrow$ true\; 
    } \label{alg:rel_end}
  }

  \tcc{RTT-agnostic CWND increase}
  \SetKwProg{Fn}{}{}{}\SetKwFunction{INC}{IncreaseCWND}%
  \Fn(){\INC{$s$, $w_{total}$}}{
    \tcc{For each ACK of subflow $s$}
    $w_s$ $\leftarrow$ $w_s$ + $1/w_{total}$ \label{alg:inc}
  }

  \tcc{Constant factor CWND decrease to ECN}
  \SetKwProg{Fn}{}{}{}\SetKwFunction{ECN}{RespondToECN}%
  \Fn(){\ECN{$s$}}{
    \tcc{For the first marked ACK of subflow $s$ per window}
    $w_s$ $\leftarrow$ $\max$ $(w_s (1 - 1/\beta)$, $cwnd_{min})$ \label{alg:ecn}
  }

  \tcc{Response to duplicate ACKs}
  \SetKwProg{Fn}{}{}{}\SetKwFunction{DEC}{DecreaseCWND}%
  \Fn(){\DEC{$s$}}{
    $w_s$ $\leftarrow$ $\max$ $(w_s/2$, $cwnd_{min})$
  }
\caption{Pseudocode of \scheme}
\label{alg:scheme}
\end{algorithm}

\subsection{\scheme algorithm} \label{s:algo}

We now discuss the exact algorithm of \scheme designed with the above four
observations. \scheme mainly consists of three components: (i) subflow
suppression/release, (ii) constant factor decrease of congestion window, and
(iii) RTT-agnostic congestion window increase.

The subflow suppression/release is a key mechanism that ensures graceful
coexistence between multipath and single-path flows. The second component
enables fast traffic shifting. The final part, as its name suggests, excludes
RTT measurements, without any performance penalty, from the part of increasing
\cwnd, which overall makes our algorithm simple.  Algorithm~\ref{alg:scheme}
shows the pseudocode of \scheme, that we explain next in detail.

\myparab{Subflow suppression/release (SSR).} The SSR mechanism permits detection
of cases where all subflows belonging to an \scheme flow struggle at the same
bottleneck link due to congestion. A representative example is a many-to-one
communication pattern (\eg, incast) where multiple flows (and subflows) compete
for bandwidth at a last mile hop (\ie, ToR switch). Upon detection, \scheme
transforms its flow to a single-path flow. Once congestion disappears, \scheme
converts its flow from a single-path flow to a multipath one.

Subflow suppression consists of two steps: detection and suppression. (1) At
detection step, \scheme checks whether the \cwnd of all its subflows has been
equal to a minimum window size for $\gamma$ number of consecutive RTTs
(lines~\ref{alg:supcheck_begin}-\ref{alg:supcheck_end} in
Algorithm~\ref{alg:scheme}). (2) At suppression step, if the previous detection
condition is met, \scheme deactivates all its subflows except for the initial
one by resetting \texttt{active} flag
(lines~\ref{alg:sup_begin}-\ref{alg:sup_end}). 

\scheme conducts subflow release similarly. If the initial subflow does not
receive any more marked packets for $\tau$ number of consecutive RTTs
(lines~\ref{alg:relcheck_begin}-\ref{alg:relcheck_end}), \scheme reactivates all
those inactivated subflows (lines~\ref{alg:rel_begin}-\ref{alg:rel_end}). When
releasing the subflows, \scheme sets \texttt{active} flag for each subflow.

Overall, while it is a simple heuristic, SSR ensures fairness between multipath
and single-path flows at a shared bottleneck link. It also helps to accommodate
more senders during an incast-like episode or to reduce the chance of costly
timeouts. We demonstrate SSR's efficacy in \sref{s:micro}.

\myparab{RTT-agnostic congestion window increase.} As discussed in
\sref{s:observe}, employing an ECN-based congestion control tends to equalize
RTTs in DCNs. The difference in RTTs for paths is at most $K$ packets where $K$
is a small marking threshold at switches (say, 10 packets). In addition, the RTT
mismatch problem does not exist in DCNs, either. Based on these insights, for
each non-duplicate ACK of subflow, we simply increase its \cwnd by $1/w_{total}$
(line~\ref{alg:inc} in Algorithm~\ref{alg:scheme}) where $w_{total}$ is the
total window size across all subflows. This ensures that \scheme can only
increase one segment per RTT across all subflows, preserving network fairness
with single-path flows at bottleneck links \cite{kelly, damon_cc}.

The amount of \cwnd increase of \scheme also strikes a right balance. Given an
congestion control algorithm $C$, let the amount of \cwnd increase of a subflow
per ACK be $C_{inc}$. For instance, the amount, $1/w_{total}$, is
$\textrm{\scheme}_{inc}$.

Now suppose RTT difference among all subflows is negligible. Then,
\eref{eq:mptcp} for DCM reduces to $a \approx w_{max}/w_{total}$ where $w_{max}$
is the maximum window size across all subflows. The increasing amount per ACK is
then about $w_{max}/(w_{total})^2$ which we call $\textrm{DCM}_{inc}$. In case of
XMP, \eref{eq:xmp} reduces to $\delta_s \approx w_s / w_{total}$. Note that
$\delta_s$ is the amount of \cwnd increase per RTT in XMP. Since $w_s$ is the
current window size of subflow $s$, the subflow would receive $w_s$ number of
ACKs. Thus, for every ACK, XMP increases \cwnd of a subflow by $1/w_{total}$,
which is $\textrm{XMP}_{inc}$. Putting it together, we have
\[\textrm{DCM}_{inc} \le \textrm{\scheme}_{inc} \approx \textrm{XMP}_{inc} \]
Note that if $w_{max}$ approaches $w_{total}$,
$\textrm{DCM}_{inc} \approx \textrm{\scheme}_{inc}$. Looking at these
relationships among three algorithms, the increment is comparable across all of
them, but \scheme's algorithm is much simpler than the other two.

\myparab{Constant factor decrease of congestion window.} In \scheme a subflow
responds to ECN signals once every window of data (i.e., approximately an RTT)
by reducing its \cwnd with a constant factor $\beta$, as depicted at
line~\ref{alg:ecn} of Algorithm~\ref{alg:scheme}. The parameter $\beta$ should
be determined such that a link is fully utilized. In other words, a queue should
not be completely drained due to \cwnd reduction. In \cite{xmp}, this problem of
choosing $\beta$ is formulated as follows:
\[ \frac{BDP+K}{\beta} \leq K, \]
Note $\beta \ge 2$; otherwise, it reduces \cwnd more aggressively than a
standard TCP. We choose $\beta$ using this formula. For instance, consider a DCN
where each link has 1Gbps speed and RTT is about $250\mu s$~\cite{dctcp} (\ie,
BDP is about 20 packets). If we set $K = 10$, $\beta \ge 3$. Since computing BDP
even for other link speed (\eg, 10Gbps) is easy, it is straightforward to set
$\beta$ after
$K$ is first determined.

\newcommand{\cm}{\ding{51}}%
\newcommand{\xm}{\ding{55}}%
\begin{table}[t]
\centering
\ra{1.3}
{
\begin{tabular}{@{}rrrc@{}} \toprule
 & \cwnd increase & Response to ECN & SSR \\ \midrule
\scheme & $w_s + 1/w_{total}$ & $ w_s (1 - 1/\beta)$ & \cm \\  \midrule
\multirow{2}{*}{DCM} & $w_s + \min(\frac{a}{w_{total}}, \frac{1}{w_s})$ & $w_s (1 - \alpha_s/2)$ & \multirow{2}{*}{\xm} \\
 & $a$ as in \eref{eq:mptcp} & $\alpha_s$ as in \eref{eq:dcm}  &  \\ \midrule
\multirow{2}{*}{XMP} & $w_{s} + \delta_{s}$ &  \multirow{2}{*}{$w_s (1-1/\beta)$} & \multirow{2}{*}{\xm} \\
  & $\delta_s$ as in \eref{eq:xmp} & & \\
\bottomrule
\end{tabular}
\caption{Summary on \scheme, DCM and XMP.}
\vspace{-.1in}
\label{tbl:summary}
}
\end{table}

\myparab{Summary.} \tref{tbl:summary} highlights key mechanisms of \scheme, DCM
and XMP. From the table, we see that \scheme is much simpler than other
solutions, easing the tuning of \scheme. A key differentiator is the subflow
suppression/release mechanism that mitigates the TCP incast and \prob.

\section{Evaluation}
\label{s:eval}

In this section we evaluate \scheme via extensive simulations using NS-3~\cite{Morteza}. For
comparison, we use DCTCP, DCM and XMP\footnote{We do not use any ECN-incapable
  TCP because it does not coexist with ECN-capable TCPs at all~\cite{judd}.}. We
first study how to tune the parameters of \scheme. We then examine \scheme under
a few basic scenarios. In particular, we will answer robustness of \scheme
against the TCP incast, its effectiveness to the \prob, and its speed in traffic
shifting. We finally study the overall performance of \scheme under a
large-scale fat-tree topology that represents a realistic data center network.

\myparab{Basic configuration.}  Throughout our simulations, the following
parameters are used without any change: (i) a link rate of 10Gbps, (ii) a link
delay of 2$\mu s$, (iii) an MSS of 1400 bytes, (iv) a maximum queue size of 100
packets, and (v) $\beta = 4$ for \scheme and XMP. We also tested \scheme over
1Gbps settings and observed that the trends were similar to those of 10Gbps
settings. We only show the results under the 10Gbps settings in interest of
space.

We set a default value for each of the following parameters: (i) the number of
subflows per multipath flow = 4, (ii) the minimum congestion window size,
$cwnd_{min}$ = 2 packets, and (iii) the ECN marking threshold, $K$ = 10 packets.
When necessary (\eg, for further analysis), we change their values.

\myparab{Evaluation metrics.} We have four key metrics: Jain's fairness
index~\cite{jain}, goodput, flow completion time (FCT) and job completion time
(JCT). We define JCT as a time period until all flows in a job finish their
transmission from its beginning.

\begin{figure}[t]
\centering 
\includegraphics[width=0.32\textwidth]{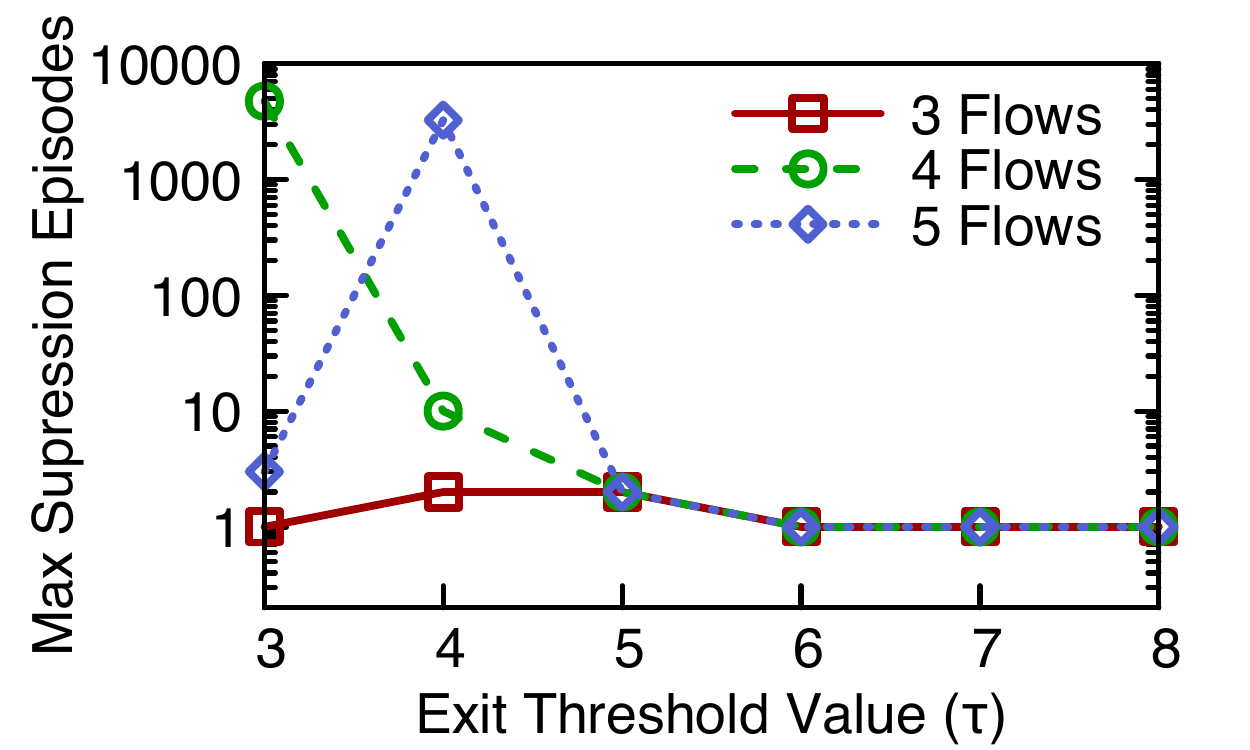}
\vspace{-.1in}
\caption{The impact of $\tau$ (i.e., the exit threshold) on the number of
  suppression episodes.}
\vspace{-.1in}
\label{fig:ssrexit} 
\end{figure}

\begin{figure*}[t]
\centering 
\subfigure[{Flow Size of 128KB}]{
\includegraphics[width=0.3\textwidth]{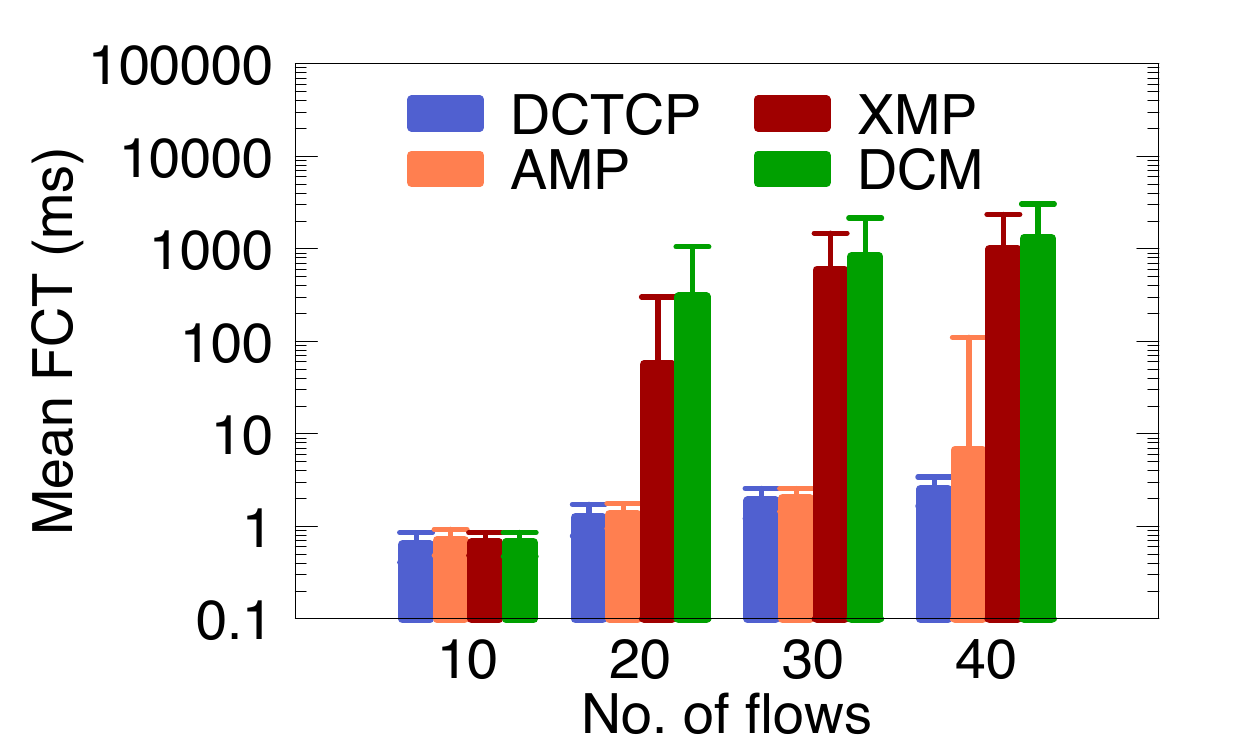}
\label{fig:incast128}}
\subfigure[{Flow Size of 256KB}]{
\includegraphics[width=0.3\textwidth]{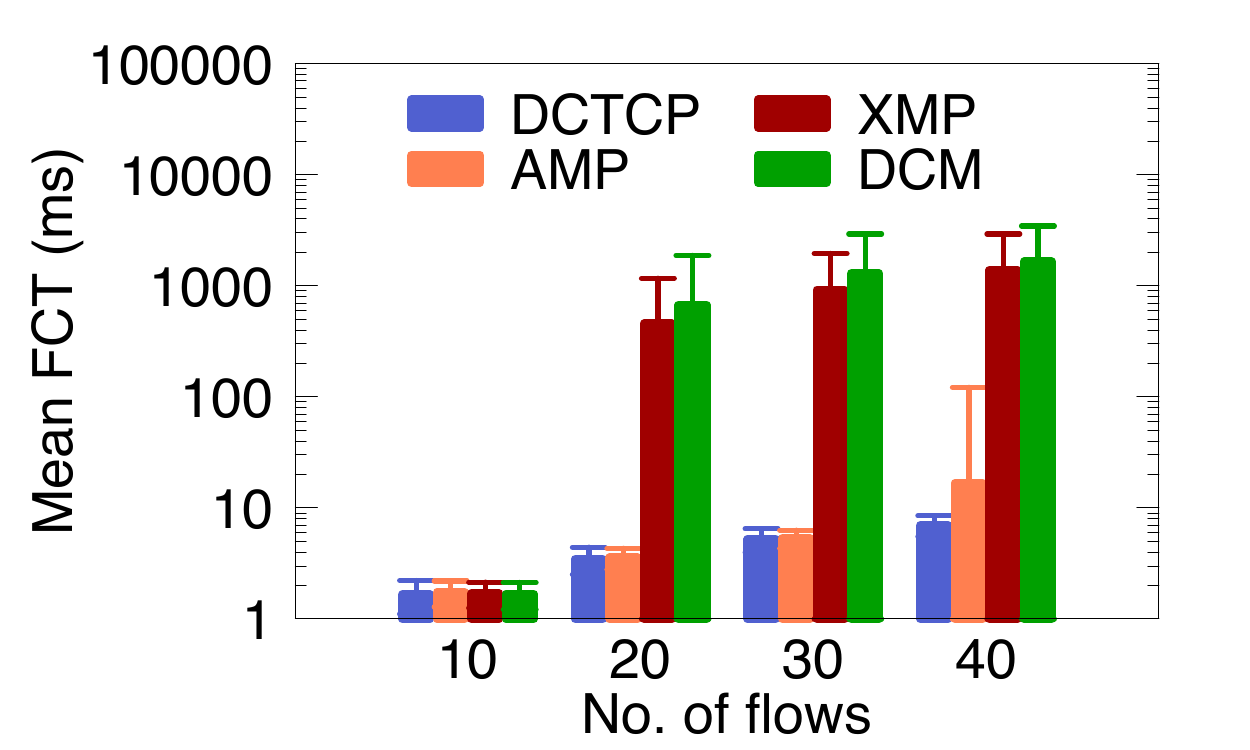}
\label{fig:incast256}} 
\subfigure[{Flow Size of 512KB}]{
\includegraphics[width=0.3\textwidth]{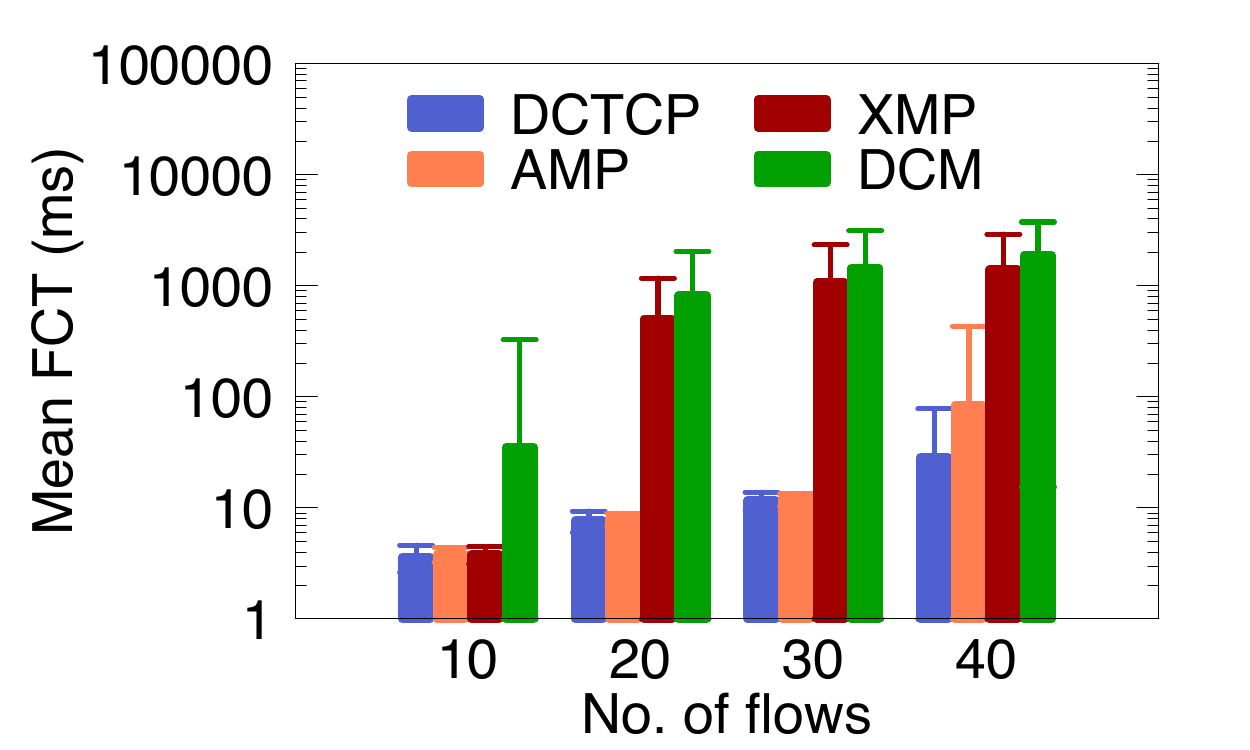}
\label{fig:incast512}} 
\vspace{-.1in}
\caption{Impact of the TCP incast on different multipath protocols. A multipath
  protocol (\scheme, XMP and DCM) is only used to transfer the incast traffic.
  A whisker bar denotes standard deviation. The y-axis is log-scaled.}
\vspace{-.1in}
\label{fig:shortflows} 
\end{figure*}

\subsection{Parameter tuning}
\label{s:ssrpara}

The subflow suppression/release (SSR) mechanism has two parameters: $\gamma$ to
begin the subflow suppression process and $\tau$ to finish it. We empirically
determine  $\gamma$ and $\tau$.

First, setting $\gamma$ is relatively easy; we test different $\gamma$ values
(1-10 RTTs) in the presence and absence of the TCP incast and \prob. If there
indeed exist the two problems in the network, it is important to begin the
suppression process early enough to alleviate their impact quickly. When
$\gamma \ge 3$ (in RTTs), \scheme reacts these problems slowly. For instance,
under the same setting for the TCP incast shown in \fref{fig:incast128kb},
average FCT of \scheme, when $\gamma = 3$, is an order of magnitude higher than
that of DCTCP. When $\gamma = 1$, there is a chance of false alarm. We find
\scheme performs best when $\gamma = 2$, which we use by default.

Second, setting $\tau$ (\ie, the exit threshold) should be more cautious. The
risk involved in selecting $\tau$ is oscillation. If $\tau$ is too small,
\scheme will repeatedly begin and end the suppression process. The frequent
oscillation may be synchronized across \scheme flows, which subsequently causes
faster queue build-up due to traffic bursts when all suspended subflows across
flows are reactivated simultaneously. This may make all incoming packets
ECN-marked, which in turn leads to the repetition of the whole suppression
process by suspending all subflows.

\begin{figure*}[t]
\centering 
\subfigure[No. of multipath flows = 1]{
\includegraphics[width=0.32\textwidth]{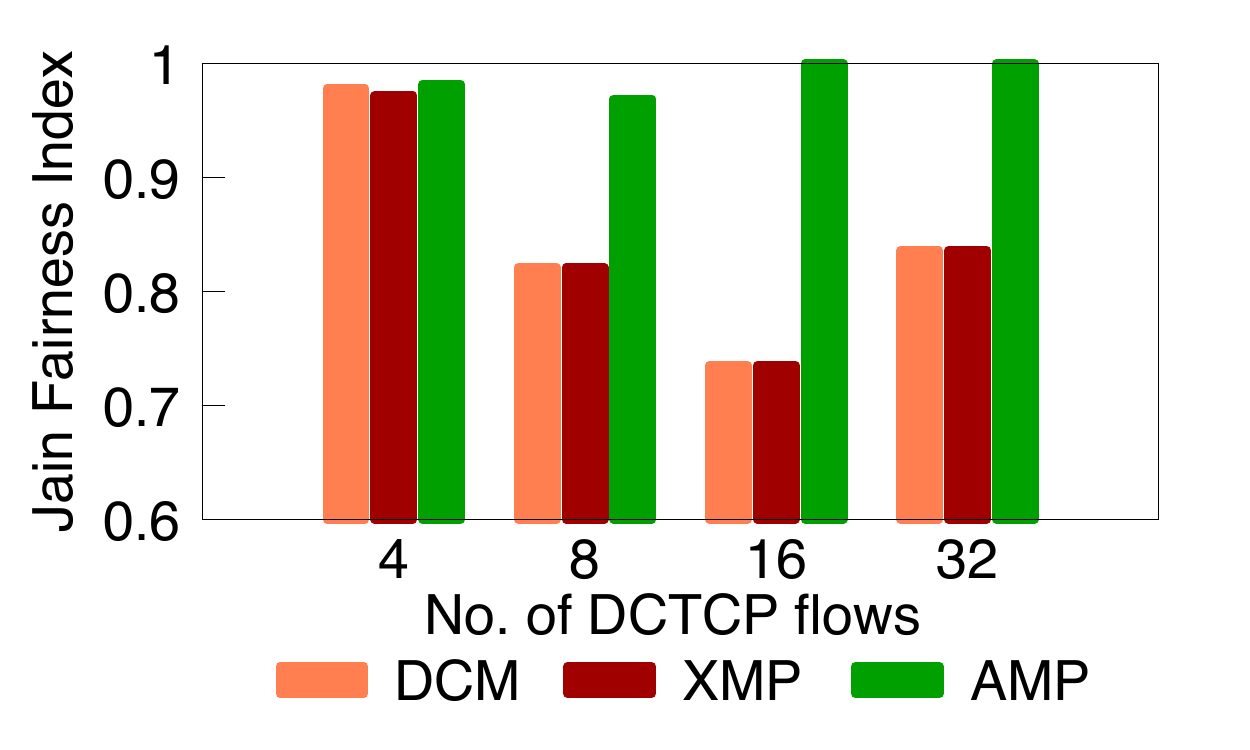} 
\label{fig:mws_mp1}} 
\hspace{-.1in}
\subfigure[No. of multipath flows = 2]{
\includegraphics[width=0.32\textwidth]{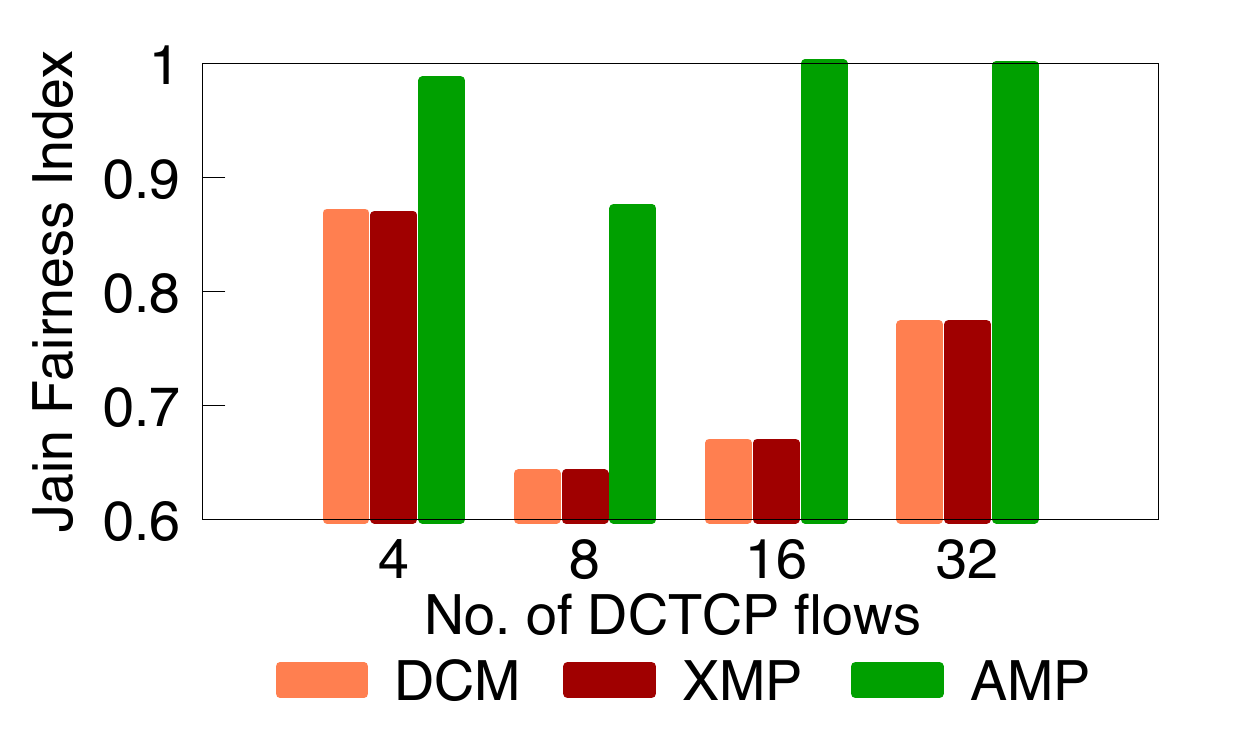} 
\label{fig:mws_mp2}} 
\hspace{-.1in}
\subfigure[No. of multipath flows = 4]{
\includegraphics[width=0.32\textwidth]{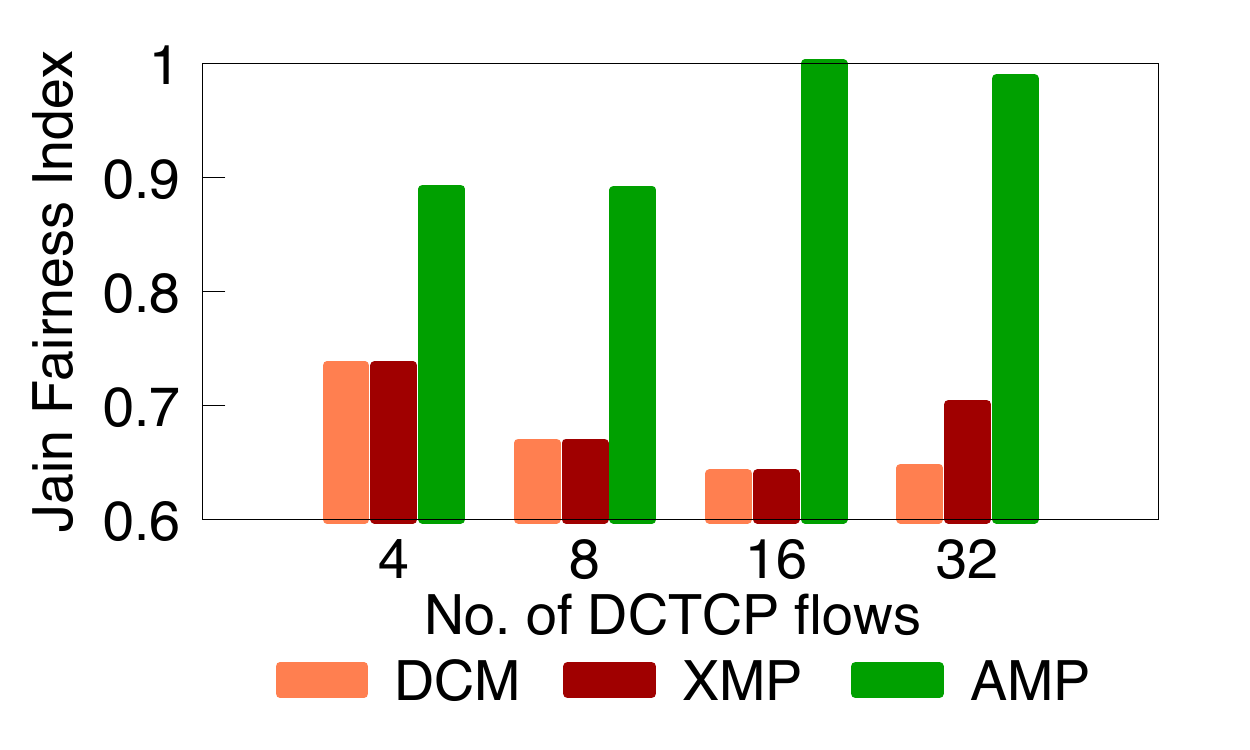} 
\label{fig:mws_mp4}}
\vspace{-.1in}
\caption{Fairness obtained when a multipath scheme (\scheme, XMP and DCM)
  competes with DCTCP flows under the \prob. Each multipath flow generates 4
  subflows. \scheme outperforms XMP and DCM.}
\vspace{-.2in}
\label{fig:mwsresult}
\end{figure*}

To find a suitable value for $\tau$, we conduct simulations while varying $\tau$
and using 3-5 \scheme flows under the topology shown in \fref{fig:startopo}.
\fref{fig:ssrexit} depicts how many suppression episodes happen across different
$\tau$ values and the number of flows. Ideally, there must be only one episode.
However, when $\tau < 6$, there are more than one episode; moreover, the
count of episodes varies a lot when the number of \scheme flows is different.
When $\tau \ge 6$, the SSR mechanism becomes stable (meaning that there is only
one episode), and the median queue length is just about 10 packets. Therefore,
we set $\tau = 8$ as default (to be conservative).

\subsection{Microbenchmarking}
\label{s:micro}

\myparab{Robustness against the TCP incast.} Multipath congestion control
mechanisms usually work poorly when they are used for traffic that is
short-lived and has a high fan-in pattern (\eg, TCP incast). To understand how
well \scheme tolerates such a traffic pattern, we use the same simulation setup
used in \sref{s:incast_motive}.
That is, there is no mix of single-path and multipath flows; we use multipath
protocols only to transfer high fan-in short-lived traffic.
This time we vary file size from 128KB to 1MB.
DCTCP is used again as baseline.

\scheme is as good as DCTCP apart from a case of 40 flows, in which \scheme
performs slightly worse than DCTCP (\fref{fig:shortflows}). However, \scheme
outperforms XMP and DCM; in most cases the average FCT of \scheme is almost 1-2
orders of magnitude shorter than that of XMP and DCM. For instance,
\fref{fig:incast128} shows that when the number of flows is 30 and flow size is
128KB, the FCT of \scheme is about 2ms and that of XMP and DCM is over 800ms. In
addition, \scheme has a narrow standard deviation in its FCT distribution, but
XMP and DCM have a large standard deviation (1-2ms for \scheme vs. 1 second for
XMP and DCM). This confirms that \scheme presents a stable FCT performance even
under various TCP incast scenarios.
Note that the y-axis of the graph is presented in log scale.

The SSR mechanism in \scheme mitigates the possibility of buffer overflow
significantly, thus that of the expensive TCP timeout.  When the number of flows
is 30 in \fref{fig:incast128}, we observe that \scheme has no timeout during the
simulation whereas XMP and DCM face up to 10 and 16 timeouts respectively (with
7 timeouts at 90th percentile for both schemes). Notably, when flow size is
smaller than 128KB (\eg, 64KB), all multipath schemes work as well as DCTCP and
there is little difference among the three multipath approaches; even the SSR
mechanism is not triggered at all as the flow size is too small. Thus, if the
flow size is at least as large as 128KB, our approach would work better than
others as we observe a similar trend for a flow size of 1MB (graph omitted).

\begin{figure}[t]
\centering 
\subfigure[One multipath flow]{
\includegraphics[width=0.235\textwidth]{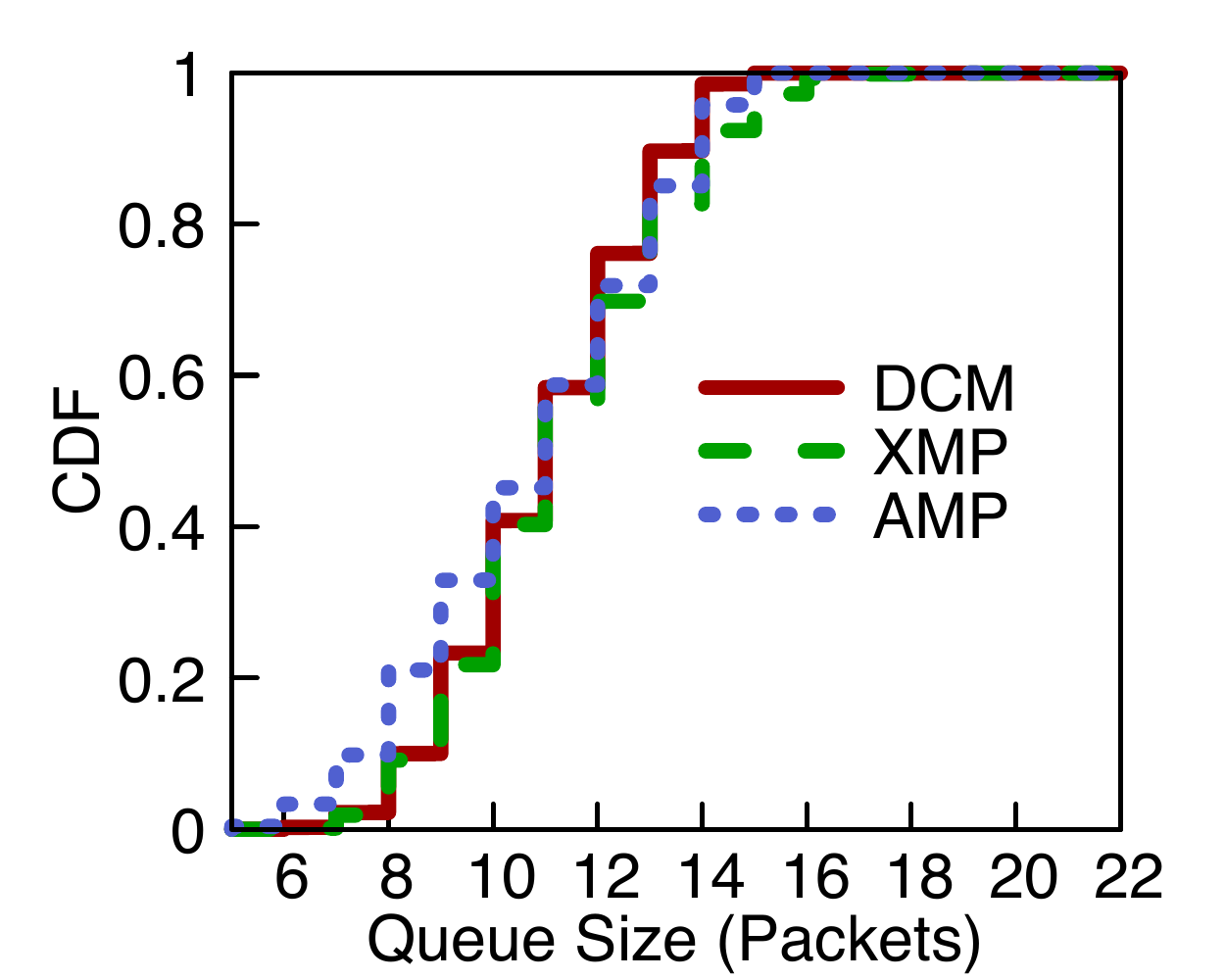}
\label{fig:mws_queue_mp1}} 
\hspace{-.23in}
\subfigure[4 multipath flows]{
\includegraphics[width=0.235\textwidth]{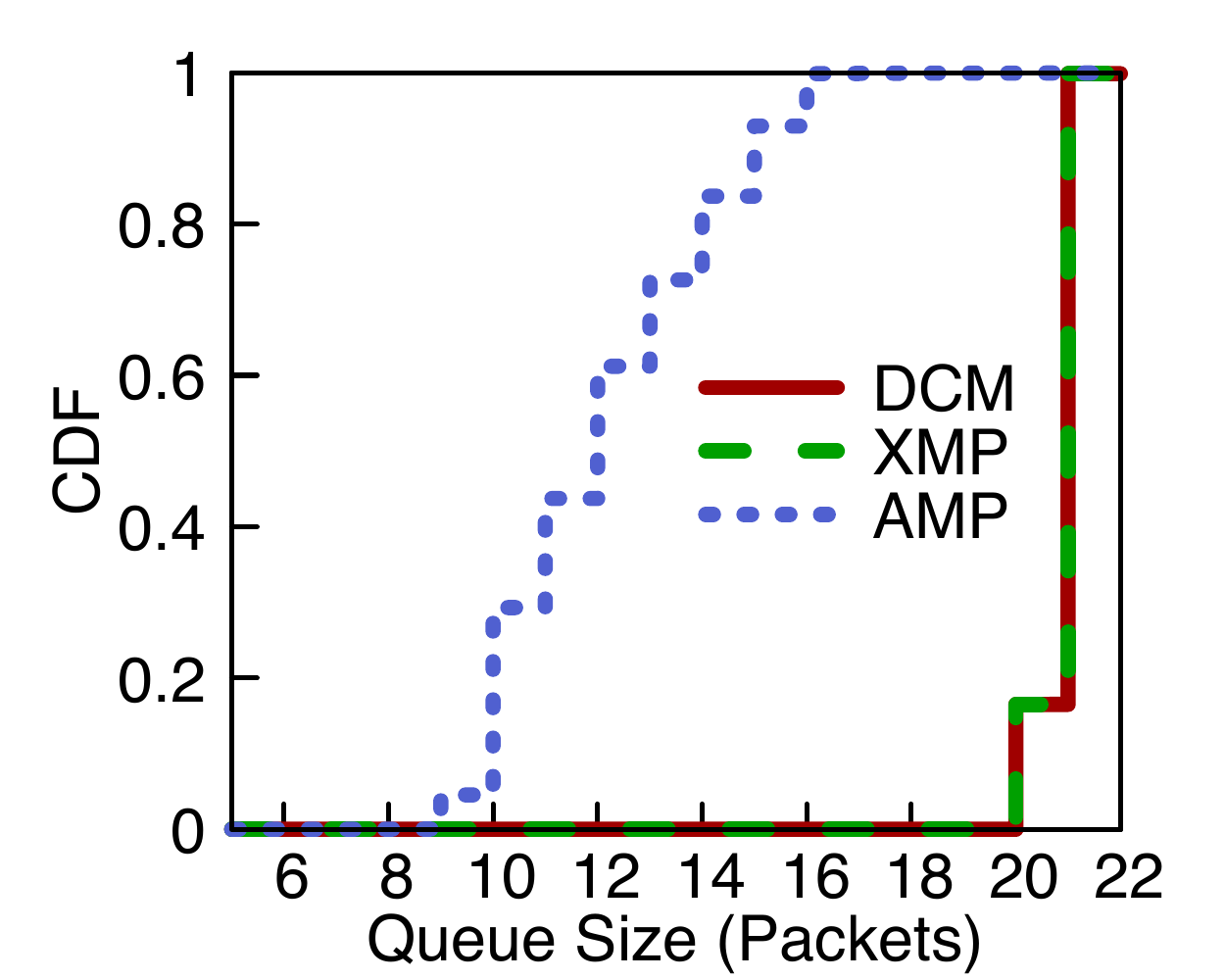}
\label{fig:mws_queue_mp4}}
\vspace{-.1in}
\caption{Queue length distribution. (a) no syndrome: 4 DCTCP flows and one
  multipath flow. (b) intensive syndrome: 4 DCTCP flows and 4 multipath flows.}
\vspace{-.1in}
\label{fig:mws_queue}
\end{figure}

\myparab{Effectiveness to the \prob.} We use the topology shown in
\fref{fig:startopo} and test the impact of the syndrome on different schemes
while varying the number of DCTCP flows and multipath flows. All flows arrive
at 0~sec and end at 1~sec.

\fref{fig:mwsresult} shows that in almost all cases \scheme outperforms the
other two schemes. As the number of multipath flows increases, we find the
syndrome aggravates fairness even in the presence of a small number of DCTCP
flows (cf., two cases of 4 DCTCP flows between Figures~\ref{fig:mws_mp1} and
\ref{fig:mws_mp4}). Note that the syndrome itself is weak in some cases (\eg,
given less than 8 flows in \fref{fig:mws_mp1} and 4 flows in
\fref{fig:mws_mp2}); thus marginal difference in fairness is observed among the
three schemes.

Since the syndrome causes persistent buffer inflation, we examine queue length.
From \fref{fig:mws_queue}, we make two observations. When there is no syndrome
(\fref{fig:mws_queue_mp1}), the queue length distributions across \scheme, XMP
and DCM are similar. On the other hand, when 4 DCM and XMP flows are used
(\fref{fig:mws_queue_mp4}), the queue length is more than 20 packets (100\%
inflation at median) all the time. On the contrary, the queue length difference
of \scheme is just about 2 packets (at median, 10 packets in
\fref{fig:mws_queue_mp1} and 12 packets in \fref{fig:mws_queue_mp4}). If the
intensity of the syndrome grows, the queue length will become more inflated
accordingly. In general \scheme can mitigate the persistent buffer inflation
better than other schemes even if the syndrome is more intensive.

The SSR mechanism is key to high performance in both fairness and delay. When we
disable SSR, all three approaches present equally poor performance.

\begin{figure}[t]
\centering 
\subfigure[\scheme]{
\includegraphics[width=0.235\textwidth]{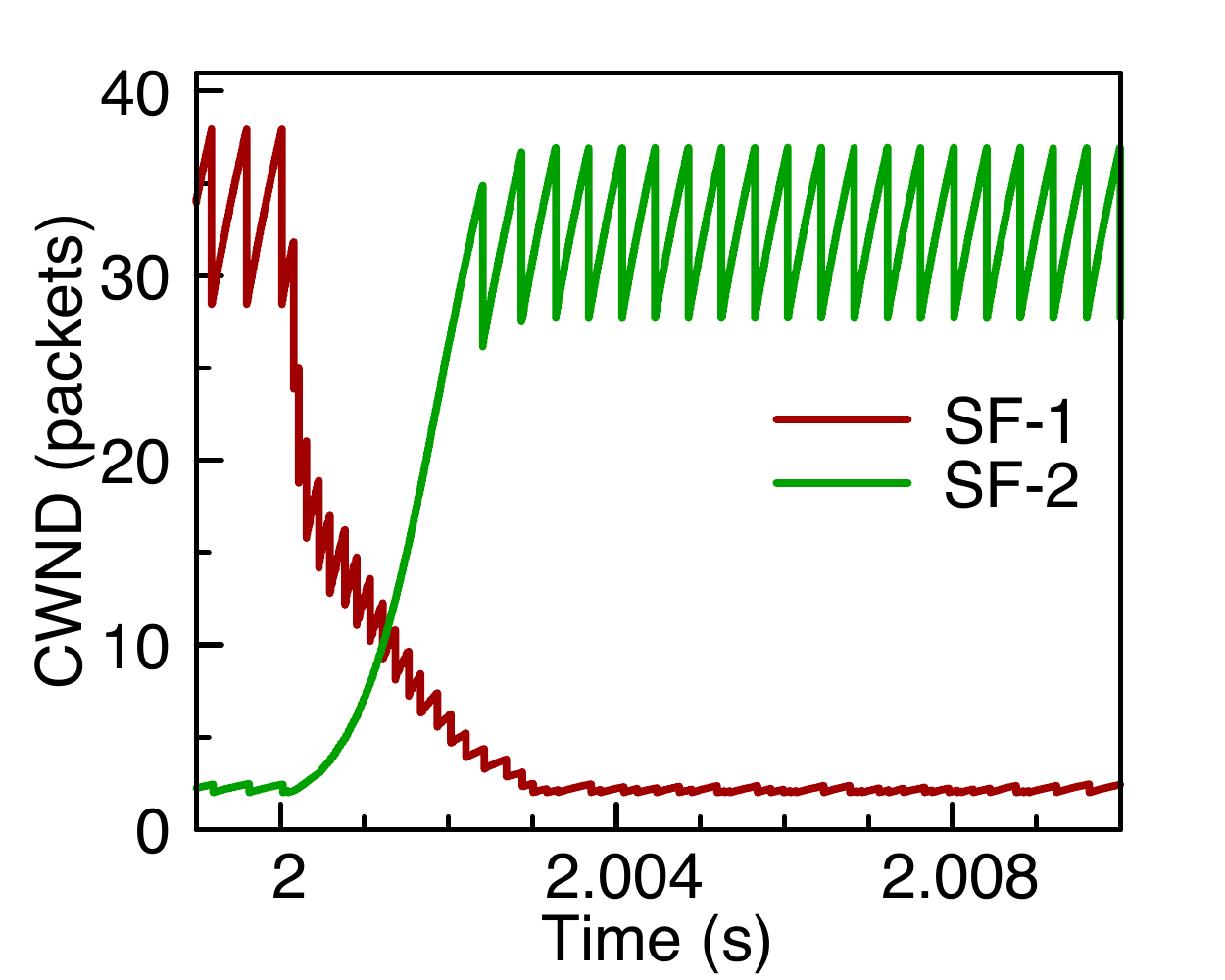}
\label{fig:amp_trash}} 
\hspace{-.23in}
\subfigure[XMP]{
\includegraphics[width=0.235\textwidth]{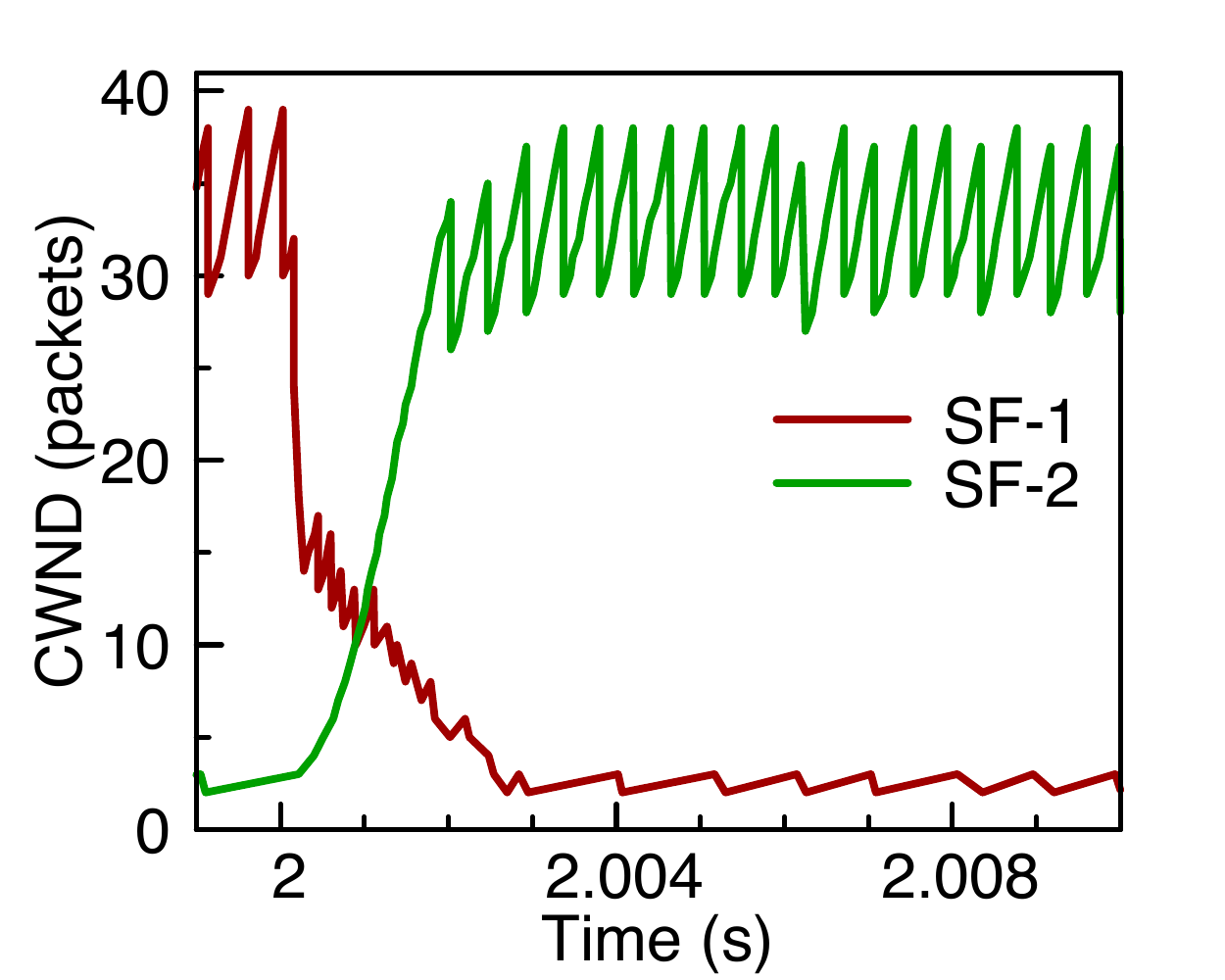}
\label{fig:xmp_trash}}
\vspace{-.1in}
\caption{Traffic shifting speed.}
\vspace{-.1in}
\label{fig:trash} 
\end{figure}

\myparab{Traffic shifting speed.} We now evaluate how quickly \scheme shifts
traffic. We run the same simulation done in \sref{s:observe} under the
setup in \fref{fig:expts}. Recall that DCM's traffic shifting time is about
7ms. \fref{fig:trash} shows that both \scheme and XMP achieve similar traffic
shifting time (about 3ms), reassuring DCTCP's slowness in shifting traffic and
suggesting that a fixed amount of congestion window reduction to ECN signals is
suitable for multipath congestion control mechanisms.

\subsection{Large-scale simulation}
\label{s:macro}

We now study the overall performance of \scheme with different workloads in a
realistic data center setup. As many data center networks employ a multi-rooted
tree topology~\cite{fattree,vl2, jupiter}, we use a 3-tier fat-tree topology that has
128 servers, 32 ToR, 32 aggregate and 16 core switches. ECMP routing is employed
to select a path on a per-flow basis.

\begin{figure*}[t]
\centering 
\subfigure[4 subflows]{
\includegraphics[width=0.3\textwidth]{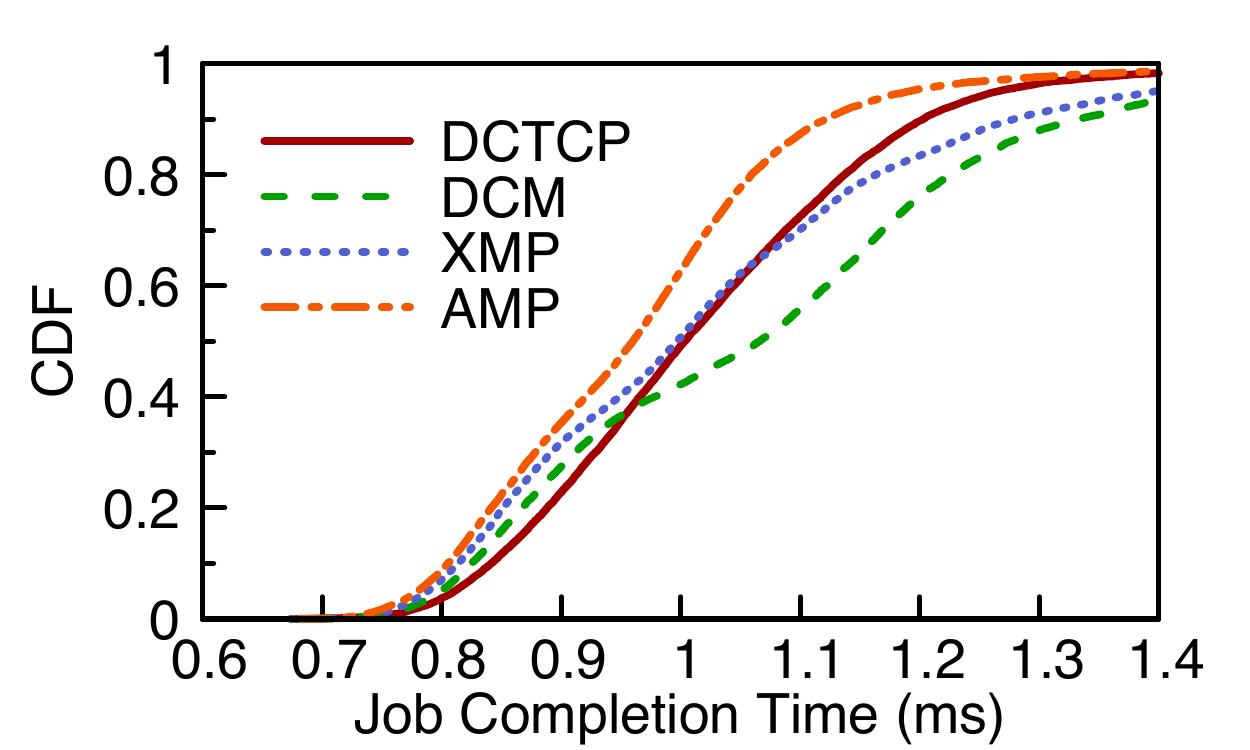}
\label{figure_in_rrss64_sf4}} 
\subfigure[6 subflows]{
\includegraphics[width=0.3\textwidth]{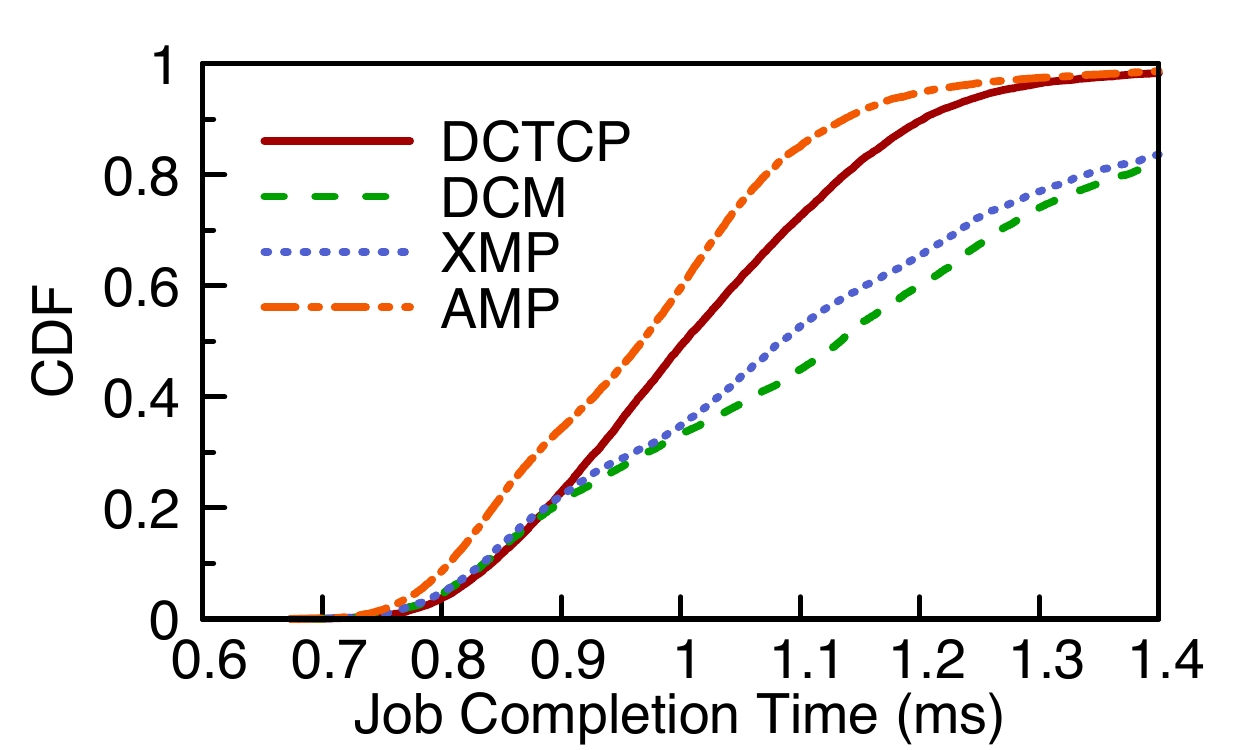}
\label{figure_in_rrss64_sf6}} 	
\subfigure[8 subflows]{
\includegraphics[width=0.3\textwidth]{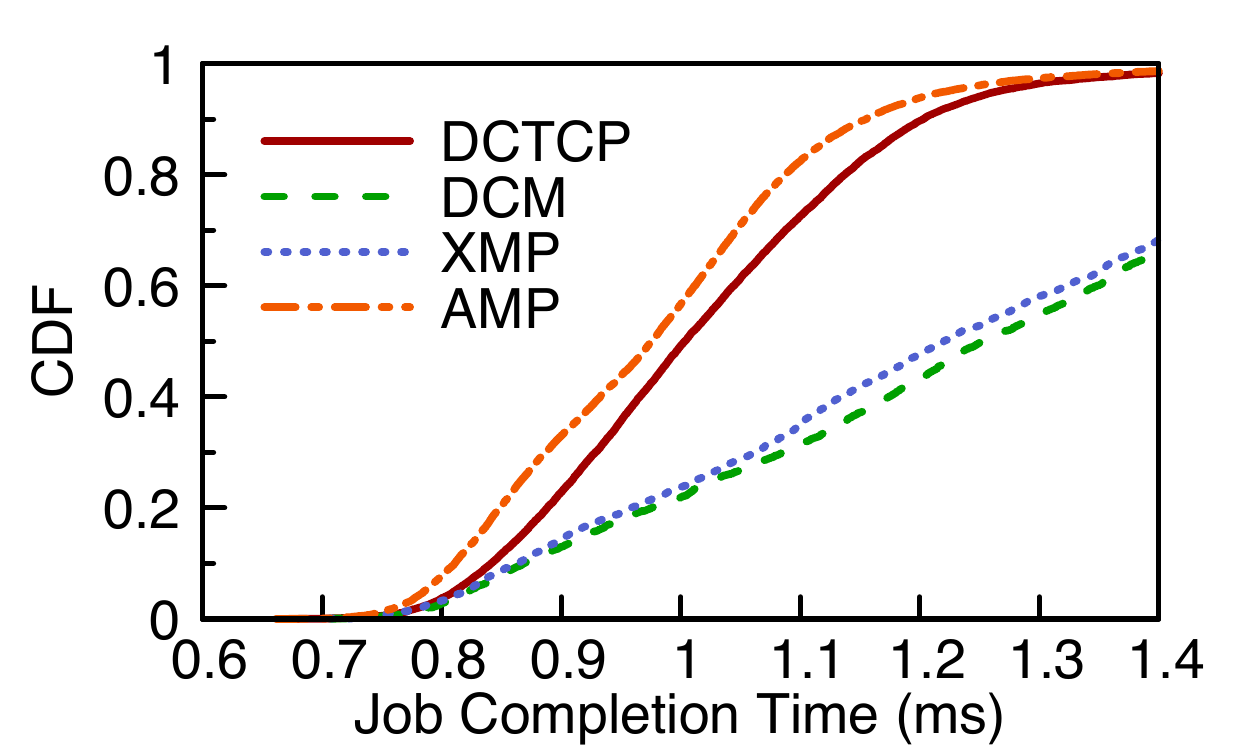}
\label{figure_in_rrss64_sf8}} 
\vspace{-.1in}
\caption{Job completion time of incast workloads. DCTCP is used for generating
  incast traffic and background traffic is generated by using \scheme, DCM, XMP
  and DCTCP (baseline) separately. A key in the legend denotes the protocol name
  used for background traffic.}
\vspace{-.1in}
\label{figure_in_rrss64} 
\end{figure*}

\begin{figure*}[t]
\centering 
\subfigure[{Short Flows}]{
\includegraphics[width=0.31\textwidth]{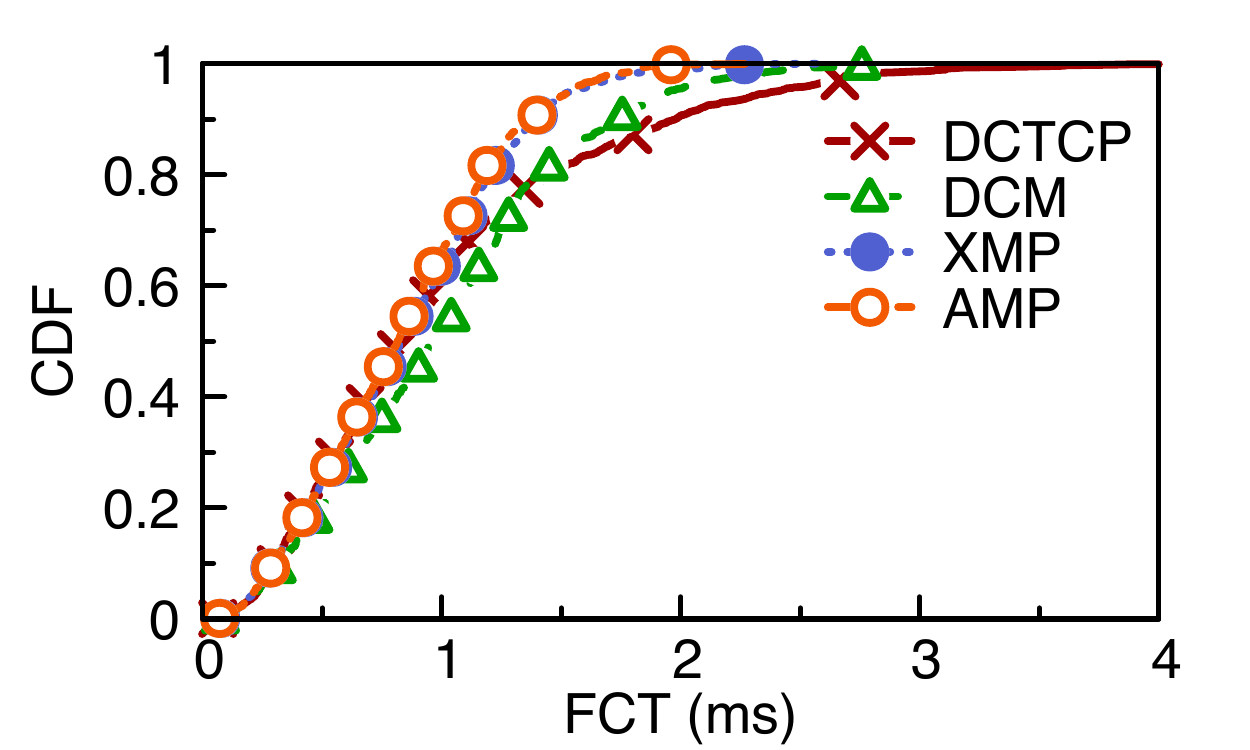} 
\label{figure_cc_l1_fct}} 
\subfigure[{Long Flows}]{
\includegraphics[width=0.31\textwidth]{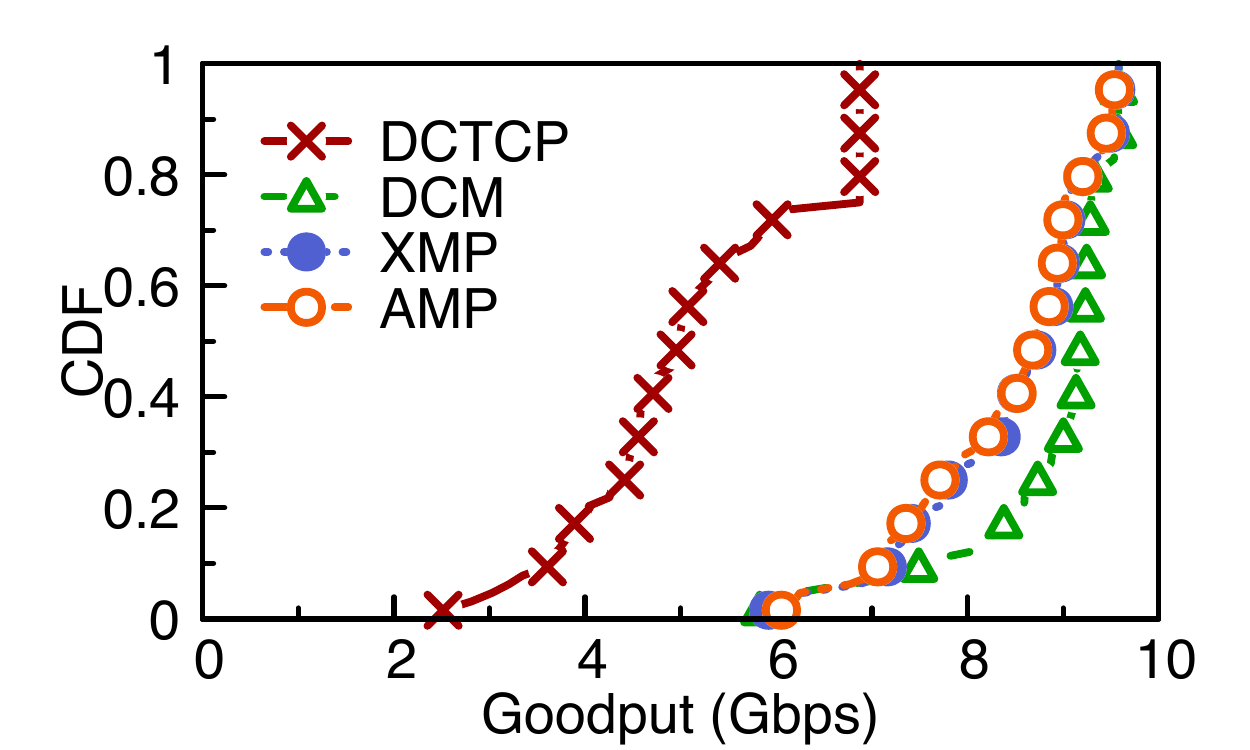}
\label{figure_cc_l1_gp}} 	
\subfigure[{Network utilization}]{
\includegraphics[width=0.31\textwidth]{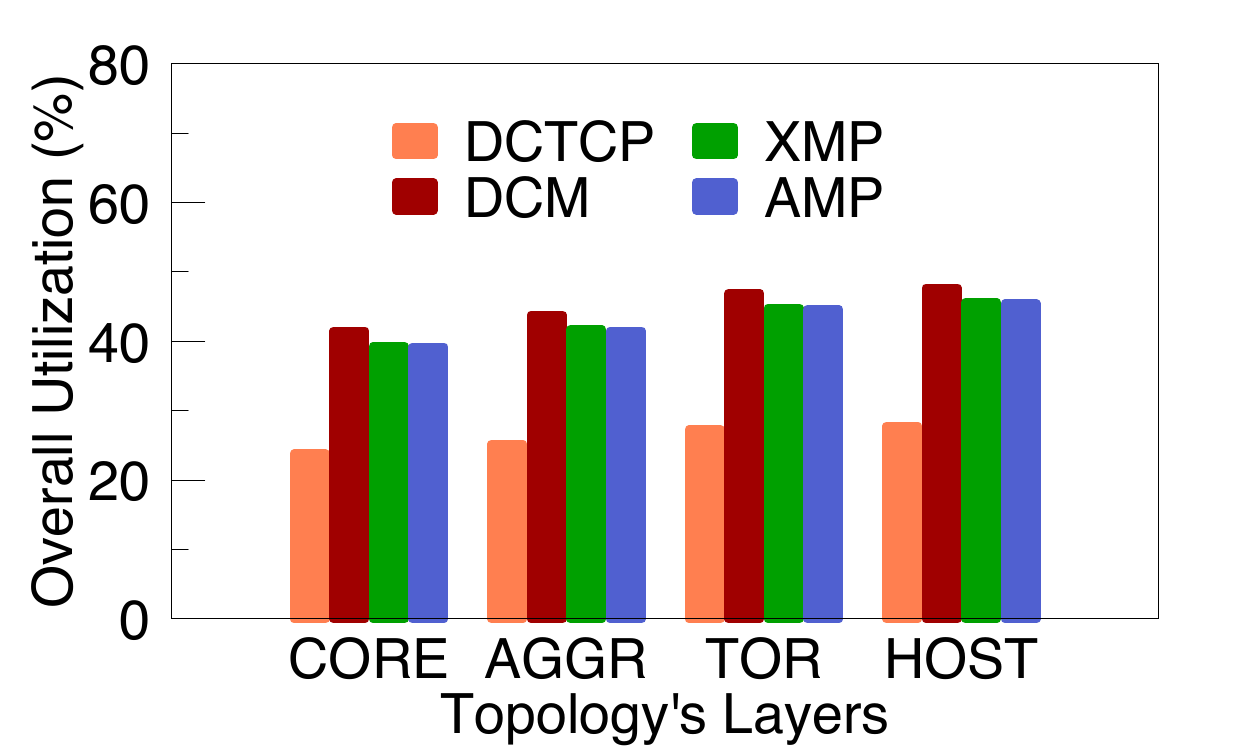} 
\label{figure_cc_l1_cu}}
\vspace{-.1in}
\caption{FCT, goodput and network utilization performance. Short flows are only
  generated by DCTCP, and each protocol in the legend is used for long flows.
  Thus, in (a), FCT is for short DCTCP flows given a different protocol for long
  flows. Similarly, (b) presents goodput of each protocol used for long flows.}
\vspace{-.1in}
\label{figure_cc_l1} 
\end{figure*}

\myparab{Incast with background traffic.} We aim to examine the performance of a
high fan-in workload (\ie, incast traffic) in the presence of background
traffic. Specifically, we use DCTCP to generate the incast traffic and a
multipath protocol for the background traffic. Note that this scenario is
different from one in \sref{s:micro} where a multipath protocol is used to
transfer the incast traffic.

\emph{Setting:} We consider a scenario where a client makes parallel reads in a
cluster filesystem in the presence of background traffic. We model this as a
unit of job: a client sends a 2KB request to 10 servers, each of which in turn
sends back a 64KB block of response data to the client. One job ends after
receiving all blocks. Thereafter a new job begins. There are 8 parallel jobs,
and clients and servers in each job are randomly selected. Each host sends a
long flow to a randomly selected host to generate traffic on background. The
flow size is determined by a Pareto distribution with shape parameter of 1.5 and
mean of 192MB. Once a long flow ends, a new one begins immediately. A simulation
continues until 1000 long flows are completed.

\emph{Results:} \fref{figure_in_rrss64} presents job completion times of short
DCTCP flows. Notice from the figure that a key in the legend is the protocol
name used for long flows. We plot, as a baseline, the case where DCTCP is also
used for long flows. Overall, we make two observations.

First, \scheme does not harm short DCTCP flows even if multipath flows use as
many as 8 subflows. The results in \fref{figure_in_rrss64} show that the \scheme
case (\ie, \scheme is used for long flows) obtains slightly better JCT
performance than the baseline case across all scenarios.

Second, more number of subflows in DCM or XMP makes the JCT of short DCTCP flows
grow quickly. When XMP is used for long flows, the 90th percentile JCT is 1.2ms
in the 4-subflow case (\fref{figure_in_rrss64_sf4}), 1.5ms in the 6-subflow
case (\fref{figure_in_rrss64_sf6}), and 1.8ms in the 8-subflow case
(\fref{figure_in_rrss64_sf8}). In contrast, the 90th percentile JCT is 1.1ms in
case where \scheme even has 8 subflows, thus reducing JCT by 0.6ms (39\%
improvement) compared to the corresponding XMP case.

\emph{Summary:} From the above observations, we conclude that our SSR mechanism
reduces buffer inflation effectively and hence makes competing short DCTCP flows
finish faster. While we recommend 4 subflows per multipath flow, \scheme may
safely support up to 8 subflows.

\myparab{General workload.} We now study interaction between short and long
flows. Our goal here is to confirm that, despite its simplicity, \scheme works
as well as other schemes and its SSR mechanism brings no harm.

\emph{Setting:}
50\% of the servers run long flows, and the remaining servers generate short
flows scheduled by a Poisson flow arrival with rate $\lambda = 256$ flows/s.
Those long flows last for 10~sec to increase chance of saturating the
network. The size for short flow is chosen between 1KB and 1MB at uniformly
random. We only present results of cases where short flows use DCTCP and long
flows use a multipath protocol because other combinations (\eg, DCTCP for long
flows and a multipath protocol for short flows) that we tested make no
significant difference in performance compared to a base case where both short
and long flows use DCTCP only.

We use permutation traffic matrix that has been used in many previous
works~\cite{hedera, xmp, mptcp, rackscale, kheirkhah2016mmptcp, phost}.  Specifically, a host
establishes at most two connections: one for receiving traffic and the other for
sending traffic. For sending traffic, the host chooses its receiver at random.

\emph{Results:} \fref{figure_cc_l1_fct} shows the FCT results of different
schemes. A key in the legend denotes a protocol used for long flows.
We observe that short DCTCP flows achieve the best FCT result when \scheme and
XMP are used for long flows. The worst FCT performance was observed when DCTCP
is used for long flows because DCTCP suffers from poor ECMP load-balancing.  In
case of long DCM flows, DCTCP flows work reasonably well and about 10\% of the
short DCTCP flows obtain longer FCTs than DCTCP flows competing with long
\scheme flows. This is because DCM's traffic shifting is slower than that of
\scheme, thus causing a high queuing delay.

The long flows of \scheme and XMP show little difference in goodput.
Interestingly, DCM achieves the best performance, which is because it trades
(queuing) delay for goodput. Again, using DCTCP for long flows yields the worst
goodput performance due to the same reason in the FCT case.

\fref{figure_cc_l1_cu} shows the mean network utilization at all layers of the
fat-tree topology. As expected,
the multipath schemes perform equally well because they balance their load among
multiple paths.

\begin{figure}[t]
\centering 
\includegraphics[width=0.32\textwidth]{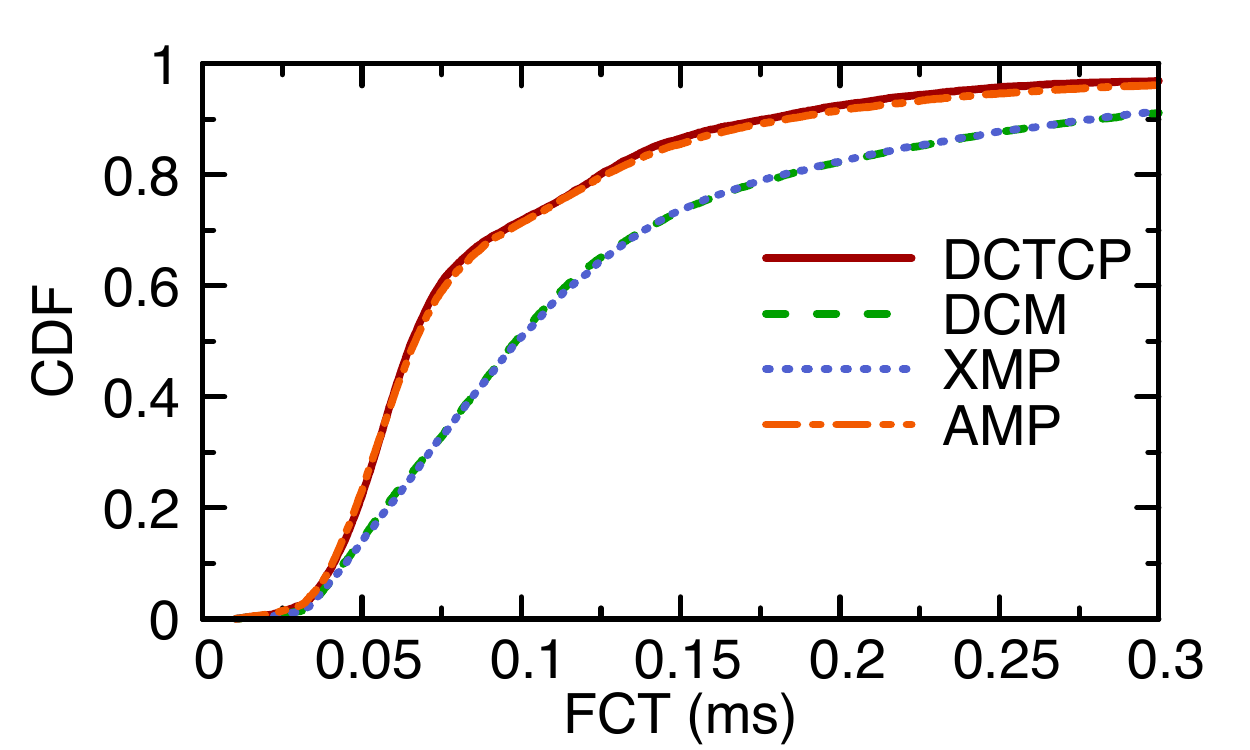}
\vspace{-.1in}
\caption{The FCT of short DCTCP flows with data mining workload used in
  pFabric~\cite{pfabric}. If a flow size $<$ 100KB, the flow is short;
  otherwise, it is considered long. The short and long flows coexist. A key in
  the legend denotes a protocol used for long flows.}
\vspace{-.1in}
\label{fig:pfabric} 
\end{figure}

We also examine those schemes in a fat-tree topology with 128 servers using a
realistic data mining workload~\cite{pfabric} and under an intensive condition
that an average inter-flow arrival time in each server is about 780$\mu s$.
\fref{fig:pfabric} depicts the FCT of short DCTCP flows when another type of
protocol is used for long flows. When AMP is used for long flows, the 90th
percentile FCT of short DCTCP flows is 0.18ms, but the corresponding FCT for XMP
and DCM is 0.27ms and 0.28ms respectively ($\sim$55\% improvement over the two
schemes). We observe a similar level of improvement at the 99th percentile:
0.78ms for \scheme, 1.35ms for XMP and 1.41ms for DCM.
Note that the goodput distribution of long flows across all schemes is almost
identical (graph omitted).

\emph{Summary:} Our results suggest that ECN alone can signal network congestion
fast enough for multipath congestion control and considering RTT as part of an
ECN-based multipath congestion control brings little benefit in the current
setting of DCNs. The results also confirm that a fixed amount of congestion
window reduction to ECN signals enables faster traffic shifting than adjusting
the window dynamically. Thus \scheme is light and as good as other solutions.

\section{Related Work}
\label{s:related}

\myparab{Pathological congestion events.} TCP incast~\cite{incast} and TCP
outcast~\cite{outcast} are well-known pathological TCP problems in DCNs.
The TCP incast is a congestion collapse incident for TCP flows that belong to
barrier-synchronized workloads with a high fan-in traffic pattern. A bursty
packet arrival overflows shallow switch buffer, leading to expensive TCP
timeouts.
In contrast, in the TCP outcast, when a few flows from one input port compete
for an output port with many flows from another input port, the few flows are
penalized more severely. These problems are fundamentally different from the
\prob studied in this work. While those problems require switch buffer overflow,
the syndrome does not;
but it results in severe unfairness between ECN-capable single-path and
multipath TCP flows.

\myparab{Multipath congestion control.} A transport layer protocol that exploits
multiple paths between source and destination has been an active area of
research~\cite{ptcp, mtcp, cmtsctp, mptcp, xmp, mmptcp-sigcomm, ps-infocom}.
MPTCP~\cite{mptcp} divides a TCP flow into multiple subflows. Since those
subflows may take different paths, MPTCP shifts traffic between its subflows to
avoid congested paths. XMP~\cite{xmp} is similar to MPTCP, but it leverages ECN
to maintain low buffer occupancy.  MMPTCP~\cite{mmptcp} uses a packet scattering
technique to improve delay performance of short flows while it acts as a regular
MPTCP for long flows. Unfortunately, these schemes fail to handle TCP incast and
\prob. On the contrary, \scheme is simpler and handles those problems better
than these schemes.

\myparab{ECN-based congestion control.} In data centers, many ECN-based
proposals adopt instant queue length based ECN. As one of the earlier works in
this category, DCTCP~\cite{dctcp} reacts to the extent of congestion estimated
from the fraction of marked packets. D2TCP~\cite{d2tcp} and L2DCT~\cite{l2dct}
build upon DCTCP; D2TCP focuses on decreasing the likelihood of missed deadlines
for TCP flows, and L2DCT aims to reduce FCT for short flows.
ECN*\cite{tuning} proposes dequeue marking to improve the performance of both
short and long flows. Since a small threshold is used in all of these schemes,
they can be prone to the \prob when ECN-capable multipath protocols are deployed
together.

\myparab{Delay-based congestion control.} Delay-based congestion control
mechanisms had continuous attention in the past for the Internet~\cite{vegas,
  comp, fast} and wireless networks~\cite{west, veno}.
Lately, refreshed interest in those mechanisms has grown in the context of
DCNs~\cite{dx, timely}.
DX~\cite{dx} and TIMELY~\cite{timely} measure queuing delays at the microsecond
granularity, and use the measurements to keep buffer occupancy low.
These approaches are single-path protocols; they may suffer from poor
load-balancing of ECMP as DCTCP does. Thus, it would be of interest to extend
AMP for delay-based schemes.

\eat{
\myparab{Virtual congestion control.}
Recent works~\cite{acdc, vcc} propose a virtual congestion control mechanism
that translates legacy TCP protocols in tenant VMs into an advanced protocol
such as DCTCP. However, these are designed only for single-path TCPs. The
interplay between a multipath TCP and the virtual mechanism is thus beyond the
scope of this paper.
}

\myparab{Scheduling, prioritization and load balancing.} A large body of work
focuses on scheduling and prioritization~\cite{pfabric, fastpass, pase, phost,
  pias, qjump}, or load balancing~\cite{conga, presto} to support low latency in
DCNs. For scheduling and prioritization, some rely on priority queuing with
multiple queues~\cite{pias, qjump}; others conduct decentralized~\cite{pfabric,
  phost} or centralized scheduling~\cite{fastpass}; one combines different
strategies adopted in prior works~\cite{pase}. Load balancing
schemes~\cite{conga, presto} break a flow down into small groups of packets,
which are in turn distributed across multiple paths. In general, these
approaches may be useful to mitigate the TCP incast and \prob.

\section{Conclusion}
\label{s:conc}

In this paper we presented that existing multipath congestion control mechanisms
fail to handle (1) the TCP incast problem that causes temporal switch buffer
overflow due to synchronized traffic arrival; and (2) the \prob that causes
persistent buffer inflation and serious unfairness. To overcome the limitation
of the existing solutions, we proposed \scheme that adaptively switches its
operation between a multiple-subflow mode and single-subflow mode. Our extensive
evaluation results showed that \scheme is simple yet effective to those problems
and in general works well, which makes deploying \scheme in data centers
attractive.

\balance
\bibliographystyle{abbrv}
\bibliography{paper}

\begin{thebibliography}{10}

\bibitem{ns3}
{Network Simulator 3}.
\newblock \url{https://www.nsnam.org/}.
\newblock Last checked: 2017-01-27.

\bibitem{fattree}
M.~Al-Fares, A.~Loukissas, and A.~Vahdat.
\newblock {A Scalable, Commodity Data Center Network Architecture}.
\newblock In {\em ACM SIGCOMM}, 2008.

\bibitem{hedera}
M.~Al-Fares, S.~Radhakrishnan, B.~Raghavan, N.~Huang, and A.~Vahdat.
\newblock {Hedera: Dynamic Flow Scheduling for Data Center Networks}.
\newblock In {\em USENIX NSDI}, 2010.

\bibitem{conga}
M.~Alizadeh, T.~Edsall, S.~Dharmapurikar, R.~Vaidyanathan, K.~Chu,
  A.~Fingerhut, V.~T. Lam, F.~Matus, R.~Pan, N.~Yadav, and G.~Varghese.
\newblock {CONGA: Distributed Congestion-aware Load Balancing for Datacenters}.
\newblock In {\em ACM SIGCOMM}, 2014.

\bibitem{dctcp}
M.~Alizadeh, A.~Greenberg, D.~A. Maltz, J.~Padhye, P.~Patel, B.~Prabhakar,
  S.~Sengupta, and M.~Sridharan.
\newblock {Data Center TCP (DCTCP)}.
\newblock In {\em ACM SIGCOMM}, 2010.

\bibitem{alizadeh11}
M.~Alizadeh, A.~Javanmard, and B.~Prabhakar.
\newblock {Analysis of DCTCP: Stability, Convergence, and Fairness}.
\newblock In {\em ACM SIGMETRICS}, 2011.

\bibitem{pfabric}
M.~Alizadeh, S.~Yang, M.~Sharif, S.~Katti, N.~McKeown, B.~Prabhakar, and
  S.~Shenker.
\newblock {pFabric: Minimal Near-optimal Datacenter Transport}.
\newblock In {\em ACM SIGCOMM}, 2013.

\bibitem{pias}
W.~Bai, L.~Chen, K.~Chen, D.~Han, C.~Tian, and H.~Wang.
\newblock {Information-agnostic Flow Scheduling for Commodity Data Centers}.
\newblock In {\em USENIX NSDI}, 2015.

\bibitem{vegas}
L.~S. Brakmo and L.~L. Peterson.
\newblock {TCP Vegas: End to end congestion avoidance on a global Internet}.
\newblock {\em IEEE Journal on selected Areas in communications},
  13(8):1465--1480, 1995.

\bibitem{xmp}
Y.~Cao, M.~Xu, X.~Fu, and E.~Dong.
\newblock Explicit multipath congestion control for data center networks.
\newblock In {\em ACM CoNEXT}, 2013.

\bibitem{fuso}
G.~Chen, Y.~Lu, Y.~Meng, B.~Li, K.~Tan, D.~Pei, P.~Cheng, L.~Luo, Y.~Xiong,
  X.~Wang, and Y.~Zhao.
\newblock {Fast and Cautious: Leveraging Multi-path Diversity for Transport
  Loss Recovery in Data Centers}.
\newblock In {\em USENIX ATC}, 2016.

\bibitem{rackscale}
P.~Costa, H.~Ballani, and D.~Narayanan.
\newblock {Rethinking the Network Stack for Rack-scale Computers}.
\newblock In {\em USENIX HoCloud}, 2014.

\bibitem{mapreduce}
J.~Dean and S.~Ghemawat.
\newblock {MapReduce: Simplified Data Processing on Large Clusters}.
\newblock In {\em USENIX OSDI}, 2004.

\bibitem{ps-infocom}
A.~Dixit, P.~Prakash, Y.~C. Hu, and R.~R. Kompella.
\newblock {On the Impact of Packet Spraying in Data Center Networks}.
\newblock In {\em IEEE INFOCOM}, 2013.

\bibitem{veno}
C.~P. Fu and S.~C. Liew.
\newblock {TCP Veno: TCP enhancement for transmission over wireless access
  networks}.
\newblock {\em IEEE Journal on selected areas in communications},
  21(2):216--228, 2003.

\bibitem{phost}
P.~X. Gao, A.~Narayan, G.~Kumar, R.~Agarwal, S.~Ratnasamy, and S.~Shenker.
\newblock {pHost: Distributed Near-optimal Datacenter Transport Over Commodity
  Network Fabric}.
\newblock In {\em ACM CoNEXT}, 2015.

\bibitem{vl2}
A.~Greenberg, J.~R. Hamilton, N.~Jain, S.~Kandula, C.~Kim, P.~Lahiri, D.~A.
  Maltz, P.~Patel, and S.~Sengupta.
\newblock {VL2: A Scalable and Flexible Data Center Network}.
\newblock In {\em ACM SIGCOMM}, 2011.

\bibitem{qjump}
M.~P. Grosvenor, M.~Schwarzkopf, I.~Gog, R.~N.~M. Watson, A.~W. Moore, S.~Hand,
  and J.~Crowcroft.
\newblock {Queues Don'T Matter when You Can JUMP Them!}
\newblock In {\em USENIX NSDI}, 2015.

\bibitem{presto}
K.~He, E.~Rozner, K.~Agarwal, W.~Felter, J.~Carter, and A.~Akella.
\newblock {Presto: Edge-based Load Balancing for Fast Datacenter Networks}.
\newblock In {\em ACM SIGCOMM}, 2015.

\bibitem{ptcp}
H.-Y. Hsieh and R.~Sivakumar.
\newblock {pTCP: An end-to-end transport layer protocol for striped
  connections}.
\newblock In {\em IEEE ICNP}, 2002.

\bibitem{cmtsctp}
J.~R. Iyengar, P.~D. Amer, and R.~Stewart.
\newblock {Concurrent multipath transfer using SCTP multihoming over
  independent end-to-end paths}.
\newblock {\em IEEE/ACM Transactions on networking (ToN)}, 14(5):951--964,
  2006.

\bibitem{jain}
R.~Jain, A.~Durresi, and G.~Babic.
\newblock {Throughput Fairness Index: An Explanation}, 1999.
\newblock ATM Forum/99-0045.

\bibitem{judd}
G.~Judd.
\newblock {Attaining the Promise and Avoiding the Pitfalls of TCP in the
  Datacenter}.
\newblock In {\em USENIX NSDI}, 2015.

\bibitem{kelly}
F.~Kelly and T.~Voice.
\newblock {Stability of End-to-end Algorithms for Joint Routing and Rate
  Control}.
\newblock {\em SIGCOMM Comput. Commun. Rev.}, 35(2):5--12, 2005.

\bibitem{kheirkhah2016mmptcp}
M.~Kheirkhah.
\newblock {\em {MMPTCP: A Novel Transport Protocol for Data Centre Networks}}.
\newblock PhD thesis, University of Sussex, 2016.

\bibitem{Morteza}
M.~Kheirkhah, I.~Wakeman, and G.~Parisis.
\newblock {Multipath-TCP in ns-3}.
\newblock In {\em Workshop on NS3}, 2014.

\bibitem{mmptcp-sigcomm}
M.~Kheirkhah, I.~Wakeman, and G.~Parisis.
\newblock {Short vs. Long Flows: A Battle That Both Can Win}.
\newblock In {\em ACM SIGCOMM}, 2015.

\bibitem{mmptcp}
M.~Kheirkhah, I.~Wakeman, and G.~Parisis.
\newblock {MMPTCP: A Multipath Transport Protocol for Data Centers}.
\newblock In {\em IEEE INFOCOM}, 2016.

\bibitem{dx}
C.~Lee, C.~Park, K.~Jang, S.~Moon, and D.~Han.
\newblock Accurate latency-based congestion feedback for datacenters.
\newblock In {\em USENIX ATC}, 2015.

\bibitem{tcprand}
S.~Lee, M.~Lee, D.~Lee, H.~Jung, and B.~Lee.
\newblock {TCPRand}: Randomizing {TCP} payload size for {TCP} fairness in data
  center networks.
\newblock In {\em {IEEE INFOCOM}}, 2015.

\bibitem{west}
S.~Mascolo, C.~Casetti, M.~Gerla, M.~Y. Sanadidi, and R.~Wang.
\newblock {TCP westwood: Bandwidth estimation for enhanced transport over
  wireless links}.
\newblock In {\em ACM MobiCom}, 2001.

\bibitem{timely}
R.~Mittal, N.~Dukkipati, E.~Blem, H.~Wassel, M.~Ghobadi, A.~Vahdat, Y.~Wang,
  D.~Wetherall, and D.~Zats.
\newblock {TIMELY: RTT-based Congestion Control for the Datacenter}.
\newblock In {\em ACM SIGCOMM}, 2015.

\bibitem{pase}
A.~Munir, G.~Baig, S.~M. Irteza, I.~A. Qazi, A.~X. Liu, and F.~R. Dogar.
\newblock {Friends, Not Foes: Synthesizing Existing Transport Strategies for
  Data Center Networks}.
\newblock In {\em ACM SIGCOMM}, 2014.

\bibitem{l2dct}
A.~Munir, I.~A. Qazi, Z.~A. Uzmi, A.~Mushtaq, S.~N. Ismail, M.~S. Iqbal, and
  B.~Khan.
\newblock Minimizing flow completion times in data centers.
\newblock In {\em IEEE INFOCOM}, 2013.

\bibitem{fastpass}
J.~Perry, A.~Ousterhout, H.~Balakrishnan, D.~Shah, and H.~Fugal.
\newblock {Fastpass: A Centralized Zero-queue Datacenter Network}.
\newblock In {\em ACM SIGCOMM}, 2014.

\bibitem{outcast}
P.~Prakash, A.~Dixit, Y.~C. Hu, and R.~Kompella.
\newblock {The TCP Outcast Problem: Exposing Unfairness in Data Center
  Networks}.
\newblock In {\em USENIX NSDI}, 2012.

\bibitem{mptcp}
C.~Raiciu, S.~Barre, C.~Pluntke, A.~Greenhalgh, and M.~Wischik, D.and~Handley.
\newblock {Improving Datacenter Performance and Robustness with Multipath TCP}.
\newblock In {\em ACM SIGCOMM}, 2011.

\bibitem{jupiter}
A.~Singh, J.~Ong, A.~Agarwal, G.~Anderson, A.~Armistead, R.~Bannon, S.~Boving,
  G.~Desai, B.~Felderman, P.~Germano, et~al.
\newblock {Jupiter rising: A decade of clos topologies and centralized control
  in google's datacenter network}.
\newblock {\em ACM SIGCOMM Computer Communication Review}, 45(4):183--197,
  2015.

\bibitem{comp}
K.~Tan, J.~Song, Q.~Zhang, and M.~Sridharan.
\newblock {A compound TCP approach for high-speed and long distance networks}.
\newblock In {\em IEEE INFOCOM}, 2006.

\bibitem{d2tcp}
B.~Vamanan, J.~Hasan, and T.~Vijaykumar.
\newblock {Deadline-aware Datacenter TCP (D2TCP)}.
\newblock In {\em ACM SIGCOMM}, 2010.

\bibitem{incast}
V.~Vasudevan, A.~Phanishayee, H.~Shah, E.~Krevat, D.~G. Andersen, G.~R. Ganger,
  G.~A. Gibson, and B.~Mueller.
\newblock {Safe and Effective Fine-grained TCP Retransmissions for Datacenter
  Communication}.
\newblock In {\em ACM SIGCOMM}, 2009.

\bibitem{fast}
D.~X. Wei, C.~Jin, S.~H. Low, and S.~Hegde.
\newblock {FAST TCP: motivation, architecture, algorithms, performance}.
\newblock {\em IEEE/ACM Transactions on Networking (ToN)}, 14(6):1246--1259,
  2006.

\bibitem{damon_cc}
D.~Wischik, C.~Raiciu, A.~Greenhalgh, and M.~Handley.
\newblock {Design, Implementation and Evaluation of Congestion Control for
  Multipath TCP}.
\newblock In {\em USENIX NSDI}, 2011.

\bibitem{ictcp}
H.~Wu, Z.~Feng, C.~Guo, and Y.~Zhang.
\newblock {ICTCP: Incast congestion control for TCP in data-center networks}.
\newblock {\em IEEE/ACM Transactions on Networking (ToN)}, 21(2):345--358,
  2013.

\bibitem{tuning}
H.~Wu, J.~Ju, G.~Lu, C.~Guo, Y.~Xiong, and Y.~Zhang.
\newblock {Tuning ECN for Data Center Networks}.
\newblock In {\em ACM CoNEXT}, 2012.

\bibitem{mtcp}
M.~Zhang, J.~Lai, A.~Krishnamurthy, L.~L. Peterson, and R.~Y. Wang.
\newblock A transport layer approach for improving end-to-end performance and
  robustness using redundant paths.
\newblock In {\em USENIX ATC}, 2004.

\end{thebibliography}

\end{document}